\documentclass[12pt]{article}
\pdfoutput=1
\usepackage{bbm}
\usepackage{color}
\usepackage{authblk}
\usepackage{latexsym}
\usepackage{epsfig,amssymb,euscript, mathrsfs,cite}   
\usepackage{amsmath}
\usepackage{mathrsfs} 
\usepackage{enumerate}
\definecolor{MyDarkBlue}{rgb}{0.15,0.15,0.45}
\usepackage[linktocpage=true]{hyperref}
\hypersetup{
colorlinks=true, 
citecolor=MyDarkBlue,
linkcolor=MyDarkBlue,
urlcolor=MyDarkBlue,
pdfauthor={},
pdftitle={},
pdfsubject={hep-th}
}

\usepackage{mathtools}

\newsavebox{\ns}
\newsavebox{\dbrane}
\newsavebox{\dbshort}

\def\be{\begin{equation}}
\def\ee{\end{equation}}
\def\bea{\begin{eqnarray}}
\def\eea{\end{eqnarray}}

\newcommand{\nn}{\notag \\}

\def\eq#1 { \begin{equation} #1 \end{equation} }

\newcommand\R{\mathbb{R}}

\newcommand\D{\mathcal{D}}

\newcommand{\rb}{\bar{r}}

\newcommand{\tr}{\mathrm{tr}}

\newlength{\sswidth}

\newcommand\cN{\mathcal{N}}

\newcommand{\zb}{\bar{z}}

\numberwithin{equation}{section}       

\begin{document}

\begin{titlepage}

\vfill

\begin{flushright}
Imperial/TP/2021/JG/02\\
ICCUB-21-XXX
\end{flushright}

\vfill

\begin{center}
   \baselineskip=16pt
   {\Large\bf Marginal deformations and RG flows\\
   for type IIB S-folds}
  \vskip 1.5cm
Igal Arav$^1$, Jerome P. Gauntlett$^2$\\
Matthew M. Roberts$^2$ and Christopher Rosen$^3$\\
     \vskip .6cm     
                          \begin{small}
                                \textit{$^1$Institute for Theoretical Physics, University of Amsterdam,\\
                                Science Park 904, PO Box 94485,\\
                                1090 GL Amsterdam, The Netherlands}
        \end{small}\\
        \begin{small}\vskip .3cm
      \textit{$^2$Blackett Laboratory, 
  Imperial College\\ London, SW7 2AZ, U.K.}
        \end{small}\\
             \begin{small}\vskip .3cm
      \textit{$^3$Departament de F\'isica Qu\'antica i Astrof\'isica and Institut de Ci\`encies del Cosmos (ICC), \\
      Universitat de Barcelona, Mart\'i Franqu\`es 1, ES-08028, \\Barcelona, Spain}
        \end{small}\\
                       \end{center}
\vfill

\begin{center}
\textbf{Abstract} 
\end{center}
\begin{quote}

We construct a continuous one parameter family of $AdS_4\times S^1\times S^5$ S-fold solutions of type IIB string theory which have
nontrivial $SL(2,\mathbb{Z})$ monodromy in the $S^1$ direction. The solutions span a subset of a 
conformal manifold that contains the known $\mathcal{N}=4$ S-fold SCFT in $d=3$, and generically preserve $\mathcal{N}=2$
supersymmetry. We also construct RG flows across dimensions, from $AdS_5\times S^5$,
dual to $\mathcal{N}=4$, $d=4$ SYM compactified with a twisted spatial circle,
to various $AdS_4\times S^1\times S^5$ S-fold solutions, dual to $d=3$ SCFTs.
We construct additional flows between the $AdS_5$ dual of the Leigh-Strassler SCFT and an $\mathcal{N}=2$ S-fold 
as well as RG flows between various S-folds.

\end{quote}

\vfill

\end{titlepage}

\tableofcontents
\newpage

\section{Introduction}\label{sec:intro}

Non-geometric solutions of string and M-theory provide an interesting framework in which to explore the AdS/CFT correspondence and in, 
particular, can reveal the existence of novel classes of strongly coupled field theories.
In this paper we will be interested in S-fold solutions of type IIB string theory of the form $AdS_4\times{S^1}\times S^5$ which have non-vanishing
$SL(2,\mathbb{Z})$ monodromy along the $S^1$ direction and are dual to $d=3$ SCFTs with either $\mathcal{N}=4,2$ or 1 supersymmetry. 
Several classes of such solutions are now known \cite{Inverso:2016eet,Assel:2018vtq,Bobev:2019jbi,Guarino:2019oct,Guarino:2020gfe,Bobev:2020fon,Arav:2021tpk} but, with the exception of the $\mathcal{N}=4$ S-fold, the dual field theories have not yet been identified making this an especially interesting topic for further study. 

The S-fold solutions are first constructed as solutions of type IIB supergravity, typically by using a lower-dimensional gauged supergravity and then uplifting to $D=10$. The supergravity solutions have a non-compact spatial direction, $\mathbb{R}$, 
which is then compactified using a duality transformation $\mathcal{M}\in SL(2,\mathbb{Z})$ within the hyperbolic conjugacy class. For example, one can take $\mathcal{M}=ST^n$, 
where $S$, $T$ are the usual generators of $SL(2,\mathbb{Z})$ and $n\ge 3$ is an integer. Thus, a single S-fold supergravity solution gives rise to a family of S-fold solutions. 

The $\mathcal{N}=4$ S-fold solution of type IIB supergravity was first presented in \cite{Inverso:2016eet} and then further discussed in \cite{Assel:2018vtq}. 
An important observation \cite{Inverso:2016eet} is that the solution can be obtained as a certain limit of a class of $\mathcal{N}=4$ Janus solutions which describe 3d $\mathcal{N}=4$ superconformal interfaces of  4d  $\mathcal{N}=4$ SYM theory \cite{DHoker:2006qeo}. Using this perspective, and the results of \cite{Gaiotto:2008sa,Gaiotto:2008sd}, a conjecture for the 3d $\mathcal{N}=4$ SCFTs 
dual to the $\mathcal{N}=4$ S-folds was given in \cite{Assel:2018vtq} (see also \cite{Terashima:2011qi,Ganor:2014pha,Gang:2015wya}). 
For the S-fold solutions that are compactified using the duality transformation $\mathcal{M}=ST^n$
one considers the $T[U(N))]$ strongly coupled theory of \cite{Gaiotto:2008sd} and gauges the $U(N)\times U(N)$ global
symmetry using an $\mathcal{N}=4$ vector multiplet. In addition one should add\footnote{Adding the Chern-Simons term
is consistent with $\mathcal{N}=3$ supersymmetry but this is expected to become enhanced to $\mathcal{N}=4$ in the IR.} a Chern-Simons term at level $n$ and the dual 3d SCFT 
is obtained by taking the IR limit. Significant support for this proposal was provided in \cite{Assel:2018vtq} by calculating the free energy of the field theory
on the three sphere using localisation techniques and showing that in the large $N$ limit it is in exact agreement with the corresponding holographic calculation. For other choices of $\mathcal{M}\in SL(2,\mathbb{Z})$ used in the S-folding, it was conjectured that the dual $\mathcal{N}=4$ $d=3$ SCFTs can be obtained as the IR limit of a class of circular quiver gauge theories with links coupled to the $T[U(N))]$ theory. Additional investigations of such quiver gauge theories have been made in \cite{Garozzo:2018kra,Garozzo:2019hbf,Garozzo:2019ejm,Garozzo:2019xzi,Garozzo:2020pmz,Beratto:2020qyk}.

One of the goals of this paper is to show that the $\mathcal{N}=4$ S-fold solutions can be obtained as the endpoint of an RG flow that starts off from $AdS_5\times S^5$ in the UV. More precisely, considering $AdS_5\times S^5$ in Poincar\'e type coordinates, we compactify one of the spatial directions using an S-duality twist which persists along the RG flow down to the $AdS_4\times{S^1}\times S^5$ S-fold solution in the IR. From the boundary point of view this corresponds to taking the complex coupling constant of $\mathcal{N}=4$ SYM theory 
to have a specific spatial dependence along the preferred spatial direction, tracing out a semi-circular arc in the upper half plane. In addition, in order to preserve supersymmetry,
the holographic RG flow solution also involves deforming $\mathcal{N}=4$ SYM theory with spatially independent fermion and boson masses.
Remarkably, we are able to construct this particular RG flow solution as an exact solution of type IIB supergravity.

More generally, we will also construct a host of additional RG flow solutions
that start in a similar manner from $AdS_5\times S^5$ in the UV 
and then flow to S-fold solutions preserving $\mathcal{N}=2$ and $\mathcal{N}=1$ supersymmetry in the IR. Before expanding on these constructions below, we first highlight that we also construct a new one (real) parameter family of $AdS_4\times{S^1}\times S^5$ S-fold solutions of type IIB supergravity that contains both the $\mathcal{N}=4$ S-fold solution of \cite{Inverso:2016eet} and the $\mathcal{N}=2$ $SU(2)\times U(1)$ invariant S-fold solution of \cite{Guarino:2020gfe,Bobev:2020fon}. This new continuous family of S-fold solutions generically preserves $\mathcal{N}=2$ supersymmetry in $d=3$ and constitutes a component\footnote{On general
grounds conformal manifolds of $d=3$ theories preserving $\mathcal{N}=2$ supersymmetry should be K\"ahler \cite{Green:2010da}.} of the conformal manifold containing the $d=3$ SCFT dual to the $\mathcal{N}=4$ S-fold. 
We calculate the free energy on the three sphere holographically and show that it is constant along the conformal manifold, as expected.
Our result thus explains the fact, observed in \cite{Bobev:2020fon}, that the 
free energy for the $\mathcal{N}=2$ $SU(2)\times U(1)$ invariant S-fold solution of \cite{Guarino:2020gfe,Bobev:2020fon} is exactly the same as that of the $\mathcal{N}=4$ S-fold. Our new one-parameter family of S-folds differs from the one parameter family of $\mathcal{N}=2$ S-folds found in
\cite{Guarino:2020gfe}, and further explored in \cite{Giambrone:2021zvp}, which does not contain the $\mathcal{N}=4$ S-fold solution.

The supergravity solutions in this paper are all constructed within $D=5$ maximal gauged supergravity \cite{Gunaydin:1984qu} and then uplifted to type IIB supergravity \cite{Schwarz:1983qr,Howe:1983sra}. More precisely, we will utilise the 10-scalar truncation of \cite{Bobev:2016nua}, which keeps the metric and ten scalar fields, as well as convenient sub-truncations thereof. 
These $D=5$ models have an important global shift symmetry, $\varphi\to \varphi+constant$, where $\varphi$ is
a preferred scalar field, the $D=5$ ``dilaton"; for the $AdS_5$ vacuum solution $\varphi$ is dual to the coupling constant of $\mathcal{N}=4$ SYM.
All of the S-fold solutions that we shall focus on are of the form $AdS_4\times \mathbb{R}$, with the $D=5$ dilaton $\varphi$ being a linear function of the coordinate on the $\mathbb{R}$ factor. After uplifting the solutions to type IIB one generates a richer family of supergravity solutions using $SL(2,\mathbb{R})$ transformations and it is within this larger family of
solutions where the S-folding identifications are made, allowing one to compactify the $\mathbb{R}$ direction. Although this may seem
a somewhat roundabout procedure, it greatly facilitates the construction of the various new S-fold solutions and their associated RG flows
by just focussing on $D=5$ solutions with linear dilatons. Indeed, the linear dilaton ansatz is an example of a Q-lattice construction\footnote{See also \cite{Andrade:2013gsa} as well as \cite{Mateos:2011tv,Donos:2014eua,Jain:2014vka,Donos:2016zpf} for some similar constructions.} \cite{Donos:2013eha}
and thus implies that the RG flow equations that one needs to solve are ODEs, rather than PDEs.

The 10-scalar model contains the $\mathcal{N}=4$ S-fold of \cite{Inverso:2016eet} as well as the $SU(2)\times U(1)$ invariant 
$\mathcal{N}=2$ S-fold in the family of \cite{Guarino:2020gfe} and the $SU(3)$ invariant $\mathcal{N}=1$ S-fold in the family of \cite{Guarino:2019oct}, all of which have linear dilatons. The model also contains the infinite classes of $\mathcal{N}=1$ $AdS_4\times \mathbb{R}$ S-fold solutions with the $D=5$ dilaton a ``linear plus periodic function" of the 
coordinate on the $\mathbb{R}$ factor that were constructed in \cite{Arav:2021tpk}.
While these remarkable examples can still fruitfully be studied in $D=5$ 
they lie outside of the Q-lattice constructions which we will focus on in this paper and one instead needs to solve PDEs; we will briefly return to this class in the discussion session.

We will construct RG flow solutions in $D=5$ that approach $AdS_5$ in the UV with a linear dilaton. In addition, the fall-off of the remaining scalar fields are associated with deforming $\mathcal{N}=4$ SYM with certain spatially constant fermion and boson mass deformations that preserve 
supersymmetry. We construct RG flows that approach all of the known $AdS_4\times \mathbb{R}$ S-fold solutions with linear dilatons that preserve $\mathcal{N}=4,2,1$ supersymmetry within the 10-scalar model, including those lying on the conformal manifold. We also construct RG flows 
that start from an $\mathcal{N}=1$ $AdS_4\times \mathbb{R}$ S-fold solution in the UV and then flow to an $\mathcal{N}=2$ $AdS_4\times \mathbb{R}$ S-fold solution in the IR, as well as providing strong evidence for the existence of similar flows to the $\mathcal{N}=4$ S-fold solution in the IR.
These solutions arise as limiting cases of the RG flows moving from the $AdS_5$ vacuum in the UV to S-folds in the IR and hence also give rise to intermediate scaling behaviour.

One of the $D=5$ supergravity models we consider contains not only the $AdS_5$ vacuum solution, but also two $AdS_5$ solutions \cite{Freedman:1999gp} each
dual to the $\mathcal{N}=1$ $d=4$ SCFT found by Leigh-Strassler (LS) \cite{Leigh:1995ep}. In these models we find that limiting cases of the RG flows moving
from the $AdS_5$ vacuum in the UV to the $\mathcal{N}=2$ $AdS_4\times \mathbb{R}$ S-fold solution in the IR develop an intermediate scaling regime that is dominated by the LS fixed point. Moreover, we also show there is an RG flow that starts from the LS fixed point in the UV, suitably S-folded and
with additional spatially homogeneous relevant deformations turned on, which flows to an $\mathcal{N}=2$ S-fold solution in the IR. 
Together these solutions begin to reveal an intricate web of interconnections between conventional SCFTs and SCFTs associated with S-folds.

The plan of the rest of the paper is as follows. In section \ref{sec:2} we first analyse the spatially dependent deformations of $\mathcal{N}=4$, $d=4$ SYM theory utilised in the RG flow solutions from a field theory perspective. In particular, we consider deformations of the complex coupling constant $\tau$ as well as fermion and boson mass deformations that depend on one spatial dimension and still preserve some supersymmetry.  This analysis, utilising the coupling of $\mathcal{N}=4$ SYM to $\mathcal{N}=4$ off-shell conformal supergravity \cite{Maxfield:2016lok}, extends the analysis of \cite{Arav:2020obl}, which considered such deformations with constant $\tau$, as well as refines the results of \cite{Maxfield:2016lok}.
In section \ref{sixscalar} we recall some features of the 10-scalar truncation of maximal $D=5$ gauged supergravity \cite{Bobev:2016nua}
which we use in the remainder of the paper.
Section \ref{sec:4} introduces our ansatz for the metric and scalar fields and presents the associated BPS equations.
Some simple RG flow solutions are presented in section \ref{lin_dil_flow}, including the analytic solution, while section \ref{sec:more elaborate} contains more elaborate RG flows as well as the new one-parameter family of $\mathcal{N}=2$ S-fold solutions.
We conclude with some discussion in section \ref{sec:disc}. Appendix \ref{derivbps} shows how the BPS equations can be recast as a gradient flow and appendix \ref{specrta} summarises the linearised BPS modes for the S-fold solutions. 
Generically, solutions to the BPS equations are associated with $\mathcal{N}=1$ supersymmetry and Appendix \ref{appc} discusses when
this can be enhanced to $\mathcal{N}=2,4$.

\vskip .5 cm
{\bf Note added:} In the final write up of this paper \cite{Guarino:2021kyp} appeared, which has some overlap with this work.
In particular, some of the RG flows that we construct here are also constructed in \cite{Guarino:2021kyp} using a complementary approach within a maximally supersymmetric $D=4$ gauged supergravity.

 \section{S-fold deformations of $\mathcal{N}=4$ SYM theory}\label{sec:2}
 
 We begin by considering $\mathcal{N}=4$ SYM theory in $d=4$ flat spacetime, with coordinates
$(t, y_i)$.  We are interested in deformations of this theory which depend on one of the three spatial coordinates, $y_3$ say,
and preserve at least $\mathcal{N}=1$ supersymmetry with respect to the remaining $d=3$ spacetime dimensions. 
In addition, we would like to compactify the $y_3$ direction with an $SL(2,\mathbb{Z})$ twist.
Thus, in particular, if $\tau=\frac{\theta}{2\pi}+i\frac{4\pi}{g^2}$ is the complex coupling constant of $\mathcal{N}=4$ SYM,
we want to consider deformations $\tau=\tau(y_3)$ with the feature that 
\begin{align}\label{Sfolding4}
\tau(y_3+\Delta y_3)=\mathcal{M}\tau(y_3)\,,
 \end{align}
with $\mathcal{M}\in SL(2,\mathbb{Z})$ and preserving supersymmetry. 
The $SL(2,\mathbb{Z})$ also acts on other deformations as we discuss below.
The supergravity solutions we construct in the sequel are associated with $\mathcal{M}$ that lie in the hyperbolic conjugacy class with $Tr \mathcal{M}>2$, as we return to below.

In order to preserve supersymmetry with $\tau=\tau(y_3)$ we need to include additional deformations.
In general, the bosonic deformations of $\mathcal{N}=4$ SYM theory can be parametrised by the bosonic
auxiliary fields of off-shell $\mathcal{N}=4$ conformal supergravity
\cite{Bergshoeff:1980is,deRoo:1985np,deRoo:1984zyh}, as highlighted in \cite{Maxfield:2016lok} and further explored in \cite{Arav:2020obl}. 
These auxiliary fields transform in specific representations of the $SU(4)$ R-symmetry of undeformed $\mathcal{N}=4$ SYM. 
We will set the one form deformations, $V_{\mu}^i{}_j$, and two form deformations, $T_{\mu\nu}^{ij}$,
transforming in the ${\bf 15}$ and ${\bf 6}$, respectively, both to zero, but we allow spatial dependence in the Lorentz scalars
$E_{ij}$, transforming in the ${\bf 10}$, and $D^{ij}{}_{kl}$, transforming in the ${\bf 20}'$. 
As discussed in \cite{Maxfield:2016lok,Arav:2020obl}, these give rise to spatially dependent fermion, $M_\psi$, and boson, $M_\phi$, mass terms in 
the deformed $\mathcal{N}=4$ SYM Lagrangian of the schematic form\footnote{See \cite{Arav:2020obl} for the precise Lagrangian and for more details on the conventions we are using; note that in this section (only) we use a mostly plus convention for the metric.}:
\begin{align}
(M_\psi)_{ij}&=-\frac{1}{2}E_{ij}\,,\nn
(M_\phi)^{ij}{}_{kl}&=\frac{1}{2}D^{ij}{}_{kl}-\frac{1}{12}\delta^{[i}{}_k\delta^{j]}{}_l(\bar E^{mn}E_{mn}-\frac{\partial_\mu\tau\partial^\mu\bar\tau}{(\text{Im}\tau)^2})\,.
\end{align}
These deformations will preserve supersymmetry provided we can find solutions
to the following equations
\begin{align}\label{masteqs}
0  =& i\frac{\partial_\mu \bar \tau}{\text{Im}\tau}\gamma^\mu\epsilon_i+E_{ij}\epsilon^j\,,\nn
0  =& -\frac{1}{2}\varepsilon^{ijlm}D_\mu E_{kl}\gamma^\mu{\epsilon}_m + D^{ij}\,_{kl}\epsilon^l +\frac{1}{2}E_{kl}\bar{E}^{l[i}\epsilon^{j]}- \frac{1}{6} E_{ml}\bar{E}^{ml}\delta^{[i}_k \epsilon^{j]} -\frac{1}{6}E_{ml}\bar E^{m[i}\delta^{j]}_k\epsilon^l\,,
\nn
0  =& D_\mu\epsilon^i\,,
\end{align}
where the chiral spinor $\epsilon^i$ transforms in the ${\bf 4}$ of $SU(4)$, the antichiral spinor $\epsilon_i$ is the conjugate of $\epsilon^i$ 
and
\begin{align}
D_\mu  E_{kl}= (\partial_\mu-a_\mu )E_{kl}\,,\qquad
D_\mu\epsilon^i=(\partial_\mu-\frac{1}{2}a_\mu)\epsilon^i\,,\qquad
a_\mu=-i\frac{\partial_\mu(\tau+\bar\tau)}{4\text{Im}\tau}\,.
\end{align}

An analysis of possible supersymmetric deformations when $\tau$, $E_{ij}$ and $D^{ij}{}_{kl}$ just depend on $y_3$ 
was undertaken in \cite{Maxfield:2016lok}, but the most general result was not obtained due to an additional assumption being made, which we mention below. Rather than presenting a general analysis, it is more 
helpful for the purposes of this paper to just present the results for a further sub-class of deformations that are
associated with the supergravity models we discuss in the sequel. Specifically, we consider deformations that are invariant under a set of discrete symmetries (see (2.2), (2.3) of \cite{Bobev:2010de} and (B.2) of \cite{Bobev:2016nua})
and focus on the following non-vanishing deformations
\begin{align}\label{simpdefs}
\tau&=i\frac{4\pi}{g^2}\,,\nn
E_{ij}&=\text{diag}(m_1,m_2,m_3,m_4),\nn
D^{12}{}_{34}&=D^{34}{}_{12},\quad
D^{23}{}_{14}=D^{14}{}_{23},\quad
D^{31}{}_{24}=D^{24}{}_{31},\nn
D^{14}{}_{14}&=D^{23}{}_{23},\quad
D^{24}{}_{24}=D^{13}{}_{13},\quad
D^{34}{}_{34}=D^{12}{}_{12},
\end{align}
with $m_i, D^{ij}{}_{kl}\in \mathbb{R}$ (and we also have the trace-free condition $D^{14}{}_{14}+D^{24}{}_{24}+D^{34}{}_{34}=0$, as usual).
By suitably restricting these ten real deformations, all of which are functions of $y_3$, we can preserve $\mathcal{N}=1,2$ or 4
$d=3$ Poincar\'e supersymmetries, as we will see.

We first pause to highlight that we have set the theta angle, $\theta$, to zero in \eqref{simpdefs} and hence
$\tau(y_3)$ lies on the imaginary axis in the upper half plane. These deformations by themselves cannot be used to implement the S-folding given in \eqref{Sfolding4}. However, having carried out the supersymmetry analysis for this class of deformations we can
act with $SL(2,\mathbb{R})$ transformations to get a larger class of deformations, with $\tau$ then lying on a semi-circle in the upper half plane, where the S-folding procedure can
be carried out as in \eqref{Sfolding4}. In addition, there is also an action on $E_{ij}$ and $\epsilon^i$. We provide a few more details at the end of this section.

With this in mind, we continue to examine the supersymmetry conditions for the deformations as in \eqref{simpdefs}, with $\tau(y_3)$ purely imaginary.
We first notice that the connection $a_\mu=0$ and hence the last condition in \eqref{masteqs} is solved for
constant $\epsilon_i$.
With only $\epsilon_4\ne 0$, we find that the first condition in \eqref{masteqs} can be solved by imposing
\begin{align}\label{emmsn1}
\kappa m_4= 2(\ln g)',\qquad \gamma^3\epsilon_4=\kappa\epsilon^4\,,
\end{align}
with $\kappa=\pm1$ and 
$m_1,m_2,m_3$ arbitrary\footnote{In (4.8)-(4.10) of \cite{Maxfield:2016lok} it was implicitly assumed that the spatial dependence of $E_{ij}$ can be factored out and hence the analysis of \cite{Maxfield:2016lok} does not include the more general cases that we consider here. However, the analysis of \cite{Maxfield:2016lok} does cover the sub-class when $\ln g$ is  linear function of $y_3$ and with constant $m_i$, which is the focus in this paper.} functions of $y_3$. 
The projection on the supersymmetry parameter implies that, generically, we are preserving 1/2 of the supersymmetries parametrised by $\epsilon^4$, which corresponds to $\mathcal{N}=1$ in $d=3$.
Finally, the remaining conditions in \eqref{masteqs} are satisfied by
\begin{align}
D^{23}{}_{14}&=D^{14}{}_{23}=\kappa\frac{1}{2}m_1'\,,\nn
D^{31}{}_{24}&=D^{24}{}_{31}=\kappa\frac{1}{2}m_2'\,,\nn
D^{12}{}_{34}&=D^{34}{}_{12}=\kappa\frac{1}{2}m_3'\,,
\end{align}
as well as
\begin{align}\label{deesn1}
D^{14}{}_{14}&=D^{23}{}_{23}=\frac{1}{12}(m_2^2+m_3^2-2m_1^2)\,,\nn
D^{24}{}_{24}&=D^{13}{}_{13}=\frac{1}{12}(m_3^2+m_1^2-2m_2^2)\,,\nn
D^{34}{}_{34}&=D^{12}{}_{12}=\frac{1}{12}(m_1^2+m_2^2-2m_3^2)\,,
\end{align}
which does indeed satisfy $D^{14}{}_{14}+D^{24}{}_{24}+D^{34}{}_{34}=0$.
In the bulk of this paper we will focus on configurations with $\ln g$ linear in the $y_3$ direction. For this class we have
$m_4$ constant and non-zero, and we will also further consider deformations with constant $m_1,m_2,m_3$. However, in the discussion section we shall return to the more general class for which $\ln g$ is not linear in $y_3$, as well as allowing for spatially varying $m_1,m_2,m_3$.

We can also consider configurations preserving  $\mathcal{N}=2$ supersymmetry in
$d=3$ with, for example, $\epsilon_3,\epsilon_4\ne 0$ and $\epsilon_1=\epsilon_2=0$.
With $g'\ne0$ we now should impose the projections
\begin{align}
\gamma^3\epsilon_3=\kappa_3\epsilon^3\,,\qquad 
\gamma^4\epsilon_4=\kappa_4\epsilon^4\,,
\end{align}
with $\kappa_3,\kappa_4=\pm1$ and take
\begin{align}\label{emmsn2}
\kappa_3m_3=
\kappa_4m_4=2 (\ln g)'\,,
\end{align}
with $m_1$ an arbitrary function of $y_3$. We also take
\begin{align}
D^{23}{}_{14}&=D^{14}{}_{23}=\frac{1}{2}\kappa_4m_1'\,,\nn
D^{31}{}_{24}&=D^{24}{}_{31}=\frac{1}{2}\kappa_3m_1'\,,\nn
D^{12}{}_{34}&=D^{34}{}_{12}=\kappa_3\kappa_4 (\ln g)''\,,
\end{align}
as well as
\begin{align}\label{deesn2}
D^{14}{}_{14}&=D^{23}{}_{23}=D^{24}{}_{24}=D^{13}{}_{13}=-\frac{1}{12}m_1^2+\frac{1}{3}[(\ln g)']^2\,,\nn
D^{34}{}_{34}&=D^{12}{}_{12}=\frac{1}{6}m_1^2-\frac{2}{3}[(\ln g)']^2\,.
\end{align}
We always must have 
\begin{align}
m_1^2=m_2^2\,,
\end{align}
and, interestingly, when $m_1', m_2'\ne0$, we must also have 
\begin{align}\label{ntwocond}
m_2=\kappa_3\kappa_4m_1\,.
\end{align}

In order to preserve $\mathcal{N}=4$ supersymmetry in
$d=3$ when $g'\ne 0$ we should consider all four $\epsilon_i\ne 0$ and impose the projections
\begin{align}
\gamma^3\epsilon_i=\kappa_i\epsilon^i\,,\qquad i=1,\dots,4\,,
\end{align}
where $\kappa_i=\pm 1$ 
and choose 
\begin{align}\label{emmsn4}
\kappa_1m_1&=\kappa_2m_2=\kappa_3m_3=\kappa_4m_4=2 (\ln g)'\,.
\end{align}
When $(\ln g)''\ne 0$, we also require the additional condition on the signs of the masses
\begin{align}\label{n4cond}
\kappa_1\kappa_2\kappa_3\kappa_4=+1\,,
\end{align}
with
\begin{align}
D^{23}{}_{14}&=D^{14}{}_{23}=\kappa_1\kappa_4 (\ln g)''\nn
D^{31}{}_{24}&=D^{24}{}_{31}=\kappa_2\kappa_4 (\ln g)''\nn
D^{12}{}_{34}&=D^{34}{}_{12}=\kappa_3\kappa_4 (\ln g)''\,,
\end{align}
as well as
\begin{align}\label{deesn4}
D^{14}{}_{14}&=D^{23}{}_{23}=
D^{24}{}_{24}=D^{13}{}_{13}=
D^{34}{}_{34}=D^{12}{}_{12}=0\,.
\end{align}
Preservation of $\mathcal{N}=3$ supersymmetry in
$d=3$ is achieved with, say, $\epsilon_2,\epsilon_3,\epsilon_4\ne 0$ but this leads to the same $\mathcal{N}=4$ preserving configuration.

For the $\mathcal{N}=4$ case, it is interesting to highlight that when $\ln g$ is linear in $y_3$, we can have arbitrary choices for the signs of the masses in
\eqref{emmsn4}. However, the known $\mathcal{N}=4$ S-fold solutions with linear dilatons, and the ones we study in later sections, are all associated with fermion masses that satisfy the condition \eqref{n4cond}. This is connected with the fact that these S-fold solutions can arise as limiting
classes of $\mathcal{N}=4$ Janus solutions which have $\ln g''\ne0$. Similar comments apply to the 
known $\mathcal{N}=2$ S-fold solutions with linear dilatons, which have masses which satisfy \eqref{ntwocond}.

For later reference we record the mass deformations of $\mathcal{N}=4$ SYM theory that are associated with each of the supergravity
truncations (given in figure \ref{truncdiag} in the next section) which we use to construct RG flows in later sections. 
In the sequel we always consider $\ln g$ linear in the $y_3$ direction and with constant $m_i$.
Below we list the symmetry group $G\subset SU(4)$ preserved by the deformations, the constraints on the masses $m_i$ and the generic
amount of $d=3$ supersymmetry preserved. The specific $D^{ij}{}_{kl}$ components involved for each case can be determined from the analysis above,
\begin{align}
G\subset SU(4):&\qquad \text{fermion masses}\qquad \qquad\quad\text{$d=3$ susy}\nn 
SO(3)\times SO(3):&\qquad m_1=m_2=m_3=m_4\quad\qquad\, \mathcal{N}=4\nn
U(1)\times U(1):&\qquad m_1=m_2, \quad m_3=m_4\quad\quad\,\,\,\,\, \,\mathcal{N}=2\nn
SU(2)\times U(1):&\qquad m_1=m_2=0, \quad m_3=m_4\quad \,\,\mathcal{N}=2\nn
SU(2):&\qquad m_1=m_2=0, \quad m_3, m_4\qquad\, \mathcal{N}=1\nn
SO(3):&\qquad m_1=m_2=m_3, \quad m_4\qquad\quad\, \mathcal{N}=1
\end{align}

We now return to the issue of the S-folding procedure. We first note that from a given set of spatially dependent supersymmetric deformations,
$(\tau, E_{ij}, D^{ij}{}_{kl}, \epsilon^i)$, we can obtain a family of supersymmetric deformations by acting with $SL(2,\mathbb{R})$ transformations. The action on $\tau$ is given, as usual, by
\begin{align}
\tau\to P\tau =\frac{a\tau+b}{c\tau+d},\qquad 
P=\begin{pmatrix}
a & b\\
 c& d
\end{pmatrix}\in SL(2,\mathbb{R})\,.
\end{align}
The action leaves $D^{ij}{}_{kl}$ inert, but $E_{ij}$ and the supersymmetry parameters $\epsilon^i$ are multiplied by a phase
\begin{align}\label{phase}
\Big(\frac{|c\tau+d|}{c\tau +d}\Big)^q\,,
\end{align}
with $q=-1$ for $E_{ij}$ and $q=-1/2$ for $\epsilon^i$ (see \cite{Maxfield:2016lok}).

The S-folding procedure starts with supersymmetric deformations $(\tau, E_{ij}, D^{ij}{}_{kl}, \epsilon^i)$, depending on $y_3$ and $\tau$ purely imaginary, which are invariant under a simultaneous shift $y_3\to y_3+\Delta y_3$ combined with an
$SL(2,\mathbb{R})$ transformation given by
\begin{align}
{\mathcal S}(\mathsf{c})=\begin{pmatrix}
e^{\mathsf{c}} & 0\\
 0& e^{-\mathsf{c}}
\end{pmatrix}\,,
\end{align}
for some $\mathsf{c}\in \mathbb{R}$.
Thus, the purely imaginary $\tau(y_3)$ has the property that
\begin{align}
\tau(y_3+\Delta y_3)=\mathcal{S}(\mathsf{c})\tau\,,\qquad
\Leftrightarrow
\qquad
\ln g(y_3+\Delta y_3)= \ln g(y_3)- \mathsf{c}\,,
\end{align}
while the remaining $(E_{ij}, D^{ij}{}_{kl}, \epsilon^i)$ are invariant (since the phase in \eqref{phase} is unity). We now consider the 
transformed deformations $(\tau', E'_{ij}, D^{'ij}{}_{kl}, \epsilon'^i)\equiv P(\tau, E_{ij}, D^{ij}{}_{kl}, \epsilon^i)$ obtained by acting with
$P\in SL(2,\mathbb{R})$ which we demand has the property 
\begin{align}
\mathcal{M}=\pm P \mathcal{S}(\mathsf{c})P^{-1}\,,
\end{align} 
with $\mathcal{M}\in SL(2,\mathbb{Z})$. This is only possible for certain values of $\mathsf{c}$; for example, it can be achieved when
$\mathsf{c}=\text{arccosh}\, \frac{n}{2}$, with $n\in \mathbb{Z}$, $n\ge 3$ and taking
\begin{align}\label{genm}
{\cal M}=\begin{pmatrix}
n & 1 \\
-1 & 0
\end{pmatrix}\,,\qquad n\ge 3\,,
\end{align}
{\it c.f.} section 3.3 of \cite{Arav:2021tpk}.  
After shifting $y_3\to y_3+\Delta y_3$ it then immediately follows that $(\tau', E'_{ij}, D^{'ij}{}_{kl}, \epsilon'^i)$ will transform by the action of $\mathcal{M}\in SL(2,\mathbb{Z})$ as needed for the S-folding. In particular, $\tau'$ will transform as in \eqref{Sfolding4} as required. The different S-foldings are labelled by the conjugacy classes of
$\mathcal{M}\in SL(2,\mathbb{Z})$ and clearly all lie within the hyperbolic conjugacy class with $|Tr(\mathcal{M})|>2$.

\section{The 10-scalar model}\label{sixscalar}
The gravitational model we study couples the $D=5$ metric with ten scalar fields \cite{Bobev:2016nua}. It is a consistent truncation of maximal $D=5$ gauged supergravity 
and hence any solution can be uplifted to obtain an exact solution of type IIB supergravity in $D=10$, as described in \cite{Bobev:2016nua,Arav:2021tpk}. The ten scalars consist of
4 complex scalars, $z^A$, each of which parametrise the coset $SU(1,1)/U(1)$ and 2 real scalars, $\beta_1,\beta_2$.
The action is given by
\begin{align}\label{bulkaction}
S=\frac{1}{4\pi G_{(5)}}\int d^5 x\sqrt{|g|}\Big[-\frac{1}{4}R +3(\partial\beta_1)^2+(\partial\beta_2)^2+ \frac{1}{2}\mathcal{K}_{A\bar{B}}\partial_{\mu}z^{A}\partial^{\mu}\bar{z}^{\bar{B}} - \mathcal{P}\Big]\,,
\end{align}
and we work with a $(+----)$ signature convention. 
Here $\mathcal{K}_{A\bar{B}}=\partial_{A}\partial_{\bar B}\mathcal{K}$ where $\mathcal{K}$ is the K\"ahler potential given by
\begin{align}\label{kpot}
\mathcal{K}=-\sum_{A=1}^4\log(1-z^A\bar z^A)\,.
\end{align}
The scalar potential ${\cal P}$ can be conveniently derived from a superpotential-like quantity
\begin{align}\label{superpotlike}
\mathcal{W} \equiv ~&\frac{1}{L}e^{2\beta_1+2\beta_2}\left(1+z^1z^2+z^1z^3+z^1z^4+z^2z^3+z^2z^4+z^3z^4+z^1z^2z^3z^4\right)\nn
+ &\frac{1}{L}e^{2\beta_1-2\beta_2}\left(1-z^1z^2+z^1z^3-z^1z^4-z^2z^3+z^2z^4-z^3z^4+z^1z^2z^3z^4\right)\nn
 +&\frac{1}{L}e^{-4\beta_1}\left(1+z^1z^2-z^1z^3-z^1z^4-z^2z^3-z^2z^4+z^3z^4+z^1z^2z^3z^4\right)\,,
\end{align}
via
\begin{align}\label{peeform}
   \mathcal{P} = \frac{1}{8}e^{\mathcal{K}}\left[\frac{1}{6}\partial_{\beta_1}\mathcal{W}\partial_{\beta_1}\overline{\mathcal{W}}
     +\frac{1}{2}\partial_{\beta_2}\mathcal{W}\partial_{\beta_2}\overline{\mathcal{W}}+\mathcal{K}^{\bar{B} A}    
      \nabla_{A}\mathcal{W}\nabla_{\bar{B}}\overline{\mathcal{W}} -\frac{8}{3}\mathcal{W}\overline{\mathcal{W}}\right]\,,
 \end{align}
where $\mathcal{K}^{\bar{B}A}$ is the inverse of ${\cal K}_{A \bar B}$ and 
$ \nabla_{A}\mathcal{W}\equiv \partial_A\mathcal{W}+\partial_A \mathcal{K}\mathcal{W}$.

It is convenient to parametrise the four complex scalars in terms of eight real scalar fields via
\begin{align}\label{zedintermsofscs}
     z^1 &= \tanh \Big[ \frac{1}{2} \big( \alpha_1 + \alpha_2 + \alpha_3 + \varphi 
       -i \phi_1 -i \phi_2 -i \phi_3 +i \phi_4 \big) \Big] \,,\nn
            z^2 &= \tanh \Big[ \frac{1}{2} \big( \alpha_1 - \alpha_2 + \alpha_3 - \varphi 
       -i \phi_1 +i \phi_2 -i \phi_3 -i\phi_4 \big) \Big] \,,\nn
     z^3 &= \tanh \Big[ \frac{1}{2} \big( \alpha_1 + \alpha_2 - \alpha_3 - \varphi 
       -i\phi_1 -i\phi_2 +i \phi_3 -i \phi_4 \big) \Big] \,,\nn
     z^4 &= \tanh \Big[ \frac{1}{2} \big( \alpha_1 - \alpha_2 - \alpha_3 + \varphi 
       -i \phi_1 +i \phi_2 +i\phi_3 +i\phi_4 \big) \Big] \,.
     \end{align}
The $AdS_5$ vacuum with all scalar fields zero uplifts to the maximally supersymmetric $AdS_5\times S^5$ solution.
For this vacuum, schematically,
the 10 real scalar fields are dual to the following hermitian operators in $\mathcal{N}=4$ SYM theory:
\begin{align}\label{opfieldmapz}
\Delta=4:\qquad \qquad \varphi&\quad \leftrightarrow \quad \tr F_{\mu\nu} F^{\mu\nu}\,, \nn
\Delta=3:\qquad \qquad  \phi_i &\quad \leftrightarrow \quad \tr(\chi_i\chi_i+\text{cubic in $Z_i$})+h.c.\,, \qquad i=1, 2, 3 \,, \nn
   \phi_4&\quad \leftrightarrow \quad \tr(\lambda \lambda+\text{cubic in $Z_i$})+h.c.\,, \nn
\Delta=2:\qquad \qquad  \alpha_i &\quad \leftrightarrow \quad \tr(Z_i^2)+h.c.\,, \qquad\qquad\qquad\qquad i=1, 2, 3\,,\nn
\beta_1 &\quad \leftrightarrow \quad \tr(|Z_1|^2+|Z_2|^2 -2 |Z_3|^2 )\,,\nn
   \beta_2 &\quad \leftrightarrow \quad \tr(|Z_1|^2-|Z_2|^2)\,.
   \end{align}
The operators of $\mathcal{N}=4$, $d=4$ SYM appearing on the right hand side of \eqref{opfieldmapz} have been written in an $\mathcal{N}=1$ language, with
$Z_i$ and $\chi_i$ the bosonic and fermionic components of the associated three chiral superfields $\Phi_i$ while
$\lambda$ is the gaugino of the vector multiplet. 
A more careful analysis shows that in terms of
the $m_i$ discussed in section \ref{sec:2}, up to an overall factor, we should identify $(m_1,m_2,m_3,m_4)\leftrightarrow (\phi_1,\phi_2,\phi_3,-\phi_4)$, and note the minus sign.
We also identify $(D^{23}{}_{14},D^{31}{}_{24},D^{12}{}_{34})\leftrightarrow ( \alpha_1, \alpha_2, \alpha_3)$
as well as
$(D^{12}{}_{12}, D^{13}{}_{13}, D^{14}{}_{14})\leftrightarrow (\beta_1,\tfrac{1}{2}(\beta_2-\beta_1),-\tfrac{1}{2}(\beta_1+\beta_2))$ .

The model is invariant under $\mathbb{Z}_2\times S_4$ discrete symmetry, which leaves $\mathcal{W}$ invariant\footnote{The model is also invariant under another discrete $S_4$ symmetry, which is a remnant of the continuous $R$-symmetry and acts on the supercharges, which we give in appendix \ref{appc}.}. The 
$\mathbb{Z}_2$ is simply given by $z^A\to -z^A$
and the other transformations are explicitly written in (2.9),(2.10) of \cite{Arav:2021tpk}. 
The model is also invariant under constant shifts of the dilaton
\begin{align}\label{dilshift}
\varphi\to\varphi+c\,,
\end{align}
and this symmetry is generated by the holomorphic Killing vector
\begin{align}\label{holkv}
l =\frac{1}{2} \sum_{A=1}^4 {(-1)^{s(A)}} \left( 1- (z^A)^2 \right) \frac{\partial}{\partial z^A}\,,
\end{align}
where $s(A)=0$ for $A=1,4$ and $s(A)=1$ for $A=2,3$. If we define
\begin{equation}
\widetilde{\mathcal{K}} \equiv \mathcal{K} + \log \mathcal{W} + \log \overline{\mathcal{W}} \,,
\end{equation}
we have $l^A \partial_A \widetilde{\mathcal{K}} + l^{\bar{A}} \partial_{\bar{A}} \widetilde{\mathcal{K}} = 0$
and the corresponding moment map
$\mu=\mu(z^A,\bar z^A)$ is given by
\begin{equation}\label{mommapcon}
\mu = i l^A \partial_A \widetilde{\mathcal{K}} = \mathcal{K}^{A\bar{B}} \partial_{\bar{B}} \mu \,  \partial_{\bar{A}}\widetilde{\mathcal{K}} 
= -\frac{i}{2} \sum_{A=1}^4 (-1)^{s(A)} \frac{z^A - \bar{z}^A}{1- z^A \bar{z}^A}\,.
\end{equation}
For later use we observe that
\begin{align}\label{anotherdisc}
z^A\to \pm \bar z^A \quad\Rightarrow \quad\widetilde{\mathcal{K}}\to \widetilde{\mathcal{K}}\,,
\qquad\mu\to \mp \mu\,.
\end{align}

The 10-scalar truncation is not a truncation to a supergravity theory. However, the conditions for a solution of the 10-scalar
model to preserve supersymmetry as a solution of $D=5$ $SO(6)$ gauged supergravity were written down in 
\cite{Bobev:2016nua} and also used in \cite{Arav:2020obl}. We will use exactly the same conventions here.
Note that these supersymmetry conditions are preserved under the  $\mathbb{Z}_2\times S_4$ discrete symmetry mentioned above.

\begin{figure}[h!]
\centering
{\includegraphics[scale=0.65]{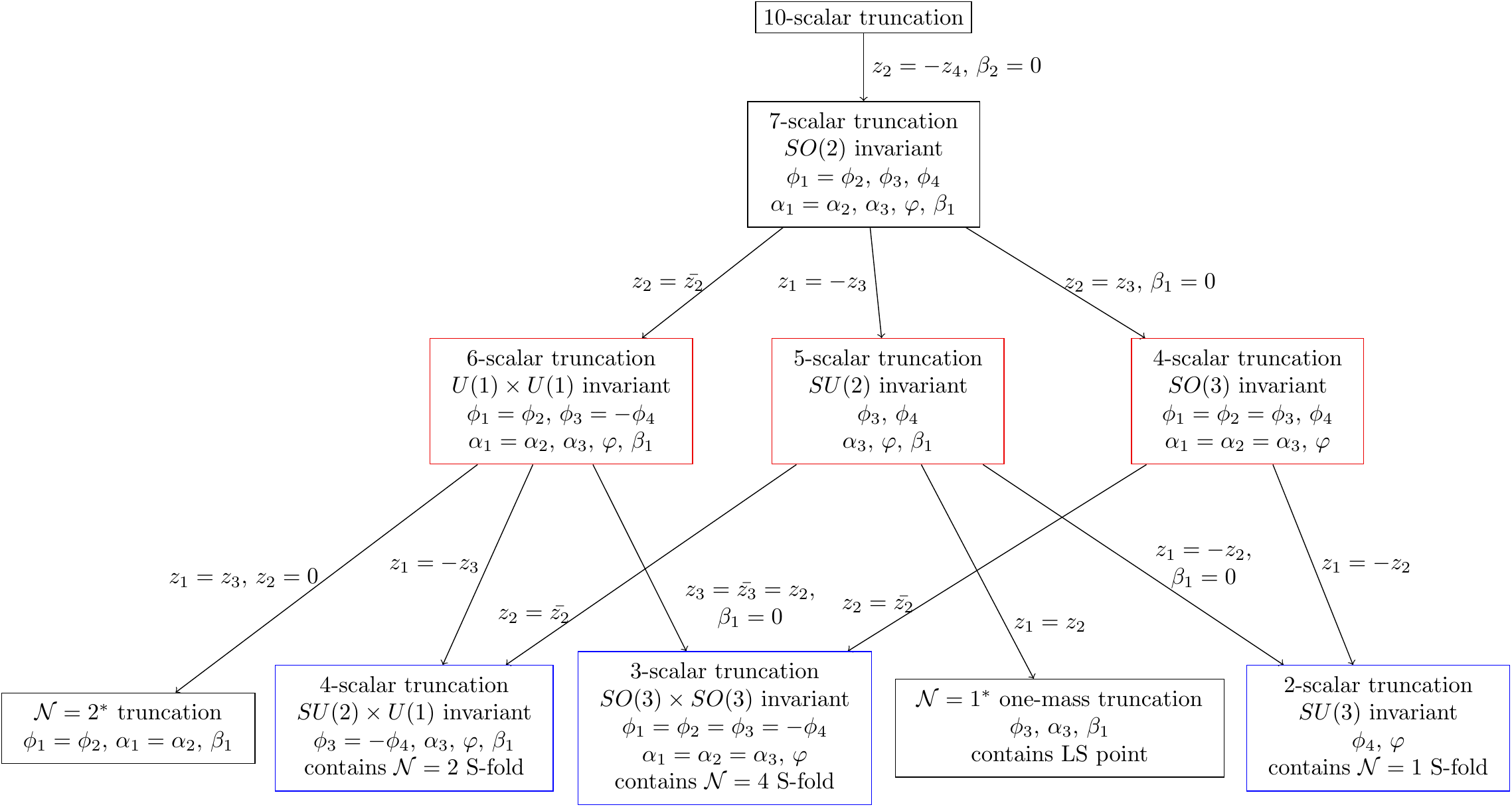}}
\caption{Various sub-truncations of the 10-scalar model in $D=5$. 
The boxes with the blue outline are truncations that contain known $AdS_4\times\mathbb{R}$ S-fold solutions with a linear dilaton.
The new one parameter family of S-folds, containing both the $\mathcal{N}=2$ and $\mathcal{N}=4$
S-folds, is contained in the $U(1)\times U(1)$ invariant 6-scalar truncation. Various RG flow solutions are discussed for the truncations in the blue and red boxes.}\label{truncdiag}
\end{figure}

Finally, there are various\footnote{There are additional consistent
truncations obtained by acting with the $\mathbb{Z}_2\times S_4$ discrete symmetry which are physically equivalent.}consistent sub-truncations of the ten-scalar model that are summarised in figure \ref{truncdiag}.
The figure also displays where various S-fold solutions,  with linear dilatons, reside. 

\section{The RG flow equations}\label{sec:4}
Within the 10-scalar model, the ansatz for the $D=5$ metric that we consider is given by
\begin{align}\label{iso21metans}
ds^2 = e^{2A}(dt^2 - dy_1^2 - dy_2^2) - e^{2V} (dx^2 + dr^2)\,,
\end{align}
with $A,V$ functions of $r$ only. This manifestly preserves $ISO(1,2)$ symmetry, associated
with the $(t,y_1,y_2)$ directions, and is also translationally invariant in the $x$ direction. 
The ansatz for the scalar fields is in essence that of a Q-lattice \cite{Donos:2013eha}: it breaks both translations in the $x$ direction
and the dilaton shift symmetry \eqref{dilshift}, but it preserves a diagonal combination. This is achieved by writing
the $D=5$ dilaton as
\begin{align}\label{dilansatz}
\varphi= kx+f(r)\,,
\end{align}
where $k$ is a constant, and taking
all other scalars to just depend on $r$.
With no essential loss of generality we will take $k>0$.
Equivalently, the $x$ dependence of the ansatz is given by
$\partial_x z^A = k l^A$, where $l^A$ are the components of the holomorphic Killing vector given in \eqref{holkv}, and $\partial_x A = \partial_x V = \partial_x \beta_i = 0$.

After substituting into the equations of motion, all of the explicit dependence on $x$ drops out and we are left with a system of ODEs for several functions of $r$.
We are interested in solutions that preserve supersymmetry. In \cite{Arav:2021tpk} BPS equations for a more general ansatz that preserves $ISO(2,1)$ symmetry were derived and
from these we can obtain BPS equations for our more restricted ansatz. As explained in appendix \ref{derivbps}, after defining
\begin{align}\label{bandpee}
b&\equiv \frac{1}{6}e^{V+\widetilde{\mathcal{K}}/2}\,,\nn
\mathcal{F} &\equiv 1 - \frac{3}{2} K^{\bar{B} A} \partial_A \widetilde{\mathcal{K}} \partial_{\bar{B}} \widetilde{\mathcal{K}} - \frac{1}{4} |\partial_{\beta_1} \mathcal{\widetilde{K}}|^2 - \frac{3}{4} |\partial_{\beta_2} \mathcal{\widetilde{K}}|^2 \,,
\end{align}
we find the BPS equations are given by
\begin{align}\label{radbpseqst}
& \partial_r A = 2 b\,,\nn
& \partial_r z^A = -3 K^{\bar{B}A} \partial_{\bar{B}}\widetilde{\mathcal{K}}b + ik l^A = -K^{\bar{B}A} \left( 3\partial_{\bar{B}} \widetilde{\mathcal{K}} b - k \partial_{\bar{B}}\mu \right) \,,\nn
& \partial_r \beta_1 = - \frac{1}{2} \partial_{\bar{\beta_1}} \widetilde{\mathcal{K}} b \,,\nn
& \partial_r \beta_2 = - \frac{3}{2} \partial_{\bar{\beta_2}} \widetilde{\mathcal{K}} b \,,\nn
&b^{-2}\partial_r b = 2 \mathcal{F}\,,
\end{align}
where the moment map $\mu $ was defined in \eqref{mommapcon}.
We emphasise that the notation for $\partial_{\bar{\beta_i}}$ is shorthand for taking $\beta_i$ to be a complex field in the superpotential such that $\mathcal{W}$ is a function of $\beta_i$ and $\overline{\mathcal{W}}$ of $\bar{\beta_i}$, then taking the appropriate derivative and finally setting $\beta_i$ to be real.
Thus, $ \partial_{\bar{\beta_1}} \mathcal{W}\equiv0$ and $\partial_{\bar{\beta_1}} \overline{\mathcal{W}}$ is given by the ordinary real derivative $\partial_{{\beta_1}} \overline{\mathcal{W}}$. We use this notation to clarify the structure of the gradient flow equations given below.
The reality of  $\beta_i$ implies constraint equations 
\begin{align}\label{consteq}
\text{Im}(\partial_{\bar{\beta_1}} \log\overline{\mathcal{W}})=0\,,
\end{align}
which is consistent with the BPS equations (see \cite{Arav:2020obl}).
Any solution to these BPS equations will generically preserve $\mathcal{N}=1$ supersymmetry in $d=3$; this can be enhanced for certain
sub-truncations as discussed in appendix \ref{appc}.

The BPS equations are invariant under the symmetry 
 \begin{align}\label{discbps}
 z^A\to -\bar z^A\,,
\end{align}
which, from \eqref{anotherdisc}, leaves the moment map invariant. Note that this is actually a symmetry of
the equations of motion of the 10-scalar model. However, it is not, in general, a symmetry of the BPS equations for general
$ISO(1,2)$ preserving configurations; it is a feature of the specific RG flow ansatz that we are considering.

 Interestingly, we can recast this as a system of gradient flow equations for 12 real functions:
\begin{align}\label{eq:10scalarGradientFlowEqst}
& e^{-\frac{3}{2}V} \partial_r z^A = - K^{\bar{B}A} \partial_{\bar{B}} P \,,\nn
& e^{-\frac{3}{2}V} \partial_r \beta_1= - \frac{1}{6} \partial_{\bar{\beta_1}} P \,,\nn
& e^{-\frac{3}{2}V} \partial_r \beta_2= - \frac{1}{2} \partial_{\bar{\beta_2}} P \,,\nn
& e^{-\frac{3}{2}V} \partial_r V = - \frac{2}{3} \partial_V P \,,\nn
& e^{-\frac{3}{2}V} \partial_r C = \frac{1}{2} P \,,
\end{align}
where $ C \equiv A + \frac{1}{2}V $ and the real function $P$ is given by
\begin{align}
& P \equiv e^{\frac{1}{2}\widetilde{\mathcal{K}} - \frac{1}{2} V} - k\mu e^{-\frac{3}{2}V} \,.
\end{align}

One might have anticipated that the BPS equations could be written as a gradient flow by arguing as follows. Associated with the ansatz for the $D=5$ fields that we are considering
we can carry out a Scherk-Schwarz dimensional reduction of the $D=5$ theory on the $x$ direction by taking the ansatz \eqref{dilansatz} with $f$ and the other scalar fields functions of the $D=4$ coordinates (and not just $r$). 
Combining this with a standard Kaluza-Klein ansatz for the $D=5$ metric then leads to a $D=4$ theory of gravity coupled to the ten scalar
fields plus an additional scalar field $V$ and a gauge-field. This $D=4$ theory has a gauge-symmetry associated with the $D=5$ shift symmetry, with gauge parameter $k$ and
$f$ is a St\"uckelberg field.
The specific radial ansatz we are considering in $D=5$ implies that in this $D=4$ theory we can consistently set the $D=4$ gauge field to zero.
The $D=5$ BPS
equations for the radial ansatz then give rise to $D=4$ BPS RG flow equations, which typically are those of a gradient flow.
Carrying out this procedure one finds the $D=4$ canonically normalised Einstein metric is given by 
\begin{align}\label{4dmet}
ds^2_{(4)}=-e^{3V}dr^2+e^{2C}(dt^2 - dy_1^2 - dy_2^2)\,,
\end{align} 
and this accounts for the factors of $e^{-\frac{3}{2}V}$ on the left hand side in
\eqref{eq:10scalarGradientFlowEqst}.
This reduced $D=4$ theory of gravity should arise as a consistent truncation of the maximal $[SO(1,1)\times SO(6)]\ltimes \mathbb{R}^{12}$ gauged supergravity 
discussed in \cite{Inverso:2016eet}. Various subtruncations of this maximal gauged supergravity have been used to study S-fold solutions in 
\cite{Guarino:2019oct,Guarino:2020gfe}.

In the rest of the paper we will analyse various specific solutions to \eqref{eq:10scalarGradientFlowEqst}. We begin with some general observations.
We first observe that
\begin{equation}\label{deepeecond}
 e^{-\frac{3}{2}V} \partial_r P = - 2 K^{\bar{B}A} \partial_{\bar{B}} P \partial_A P - \frac{1}{3} |\partial_{\beta_1} P|^2 - |\partial_{\beta_2} P|^2 - \frac{2}{3} (\partial_V P)^2 \leq 0 \,,
\end{equation}
and therefore $P$ is monotonic along the flow. 
We next notice that in any $AdS_5$ solution the ten $D=5$ scalars are constant and 
$e^A=e^V\propto r^{-1}$; while not a solution of the BPS equations \eqref{eq:10scalarGradientFlowEqst} we can approach an $AdS_5$ solution
as a limit with $P=0$. Of most interest in this paper
is the $AdS_5$ vacuum dual to $\mathcal{N}=4$ SYM theory. We will construct flows that approach this $AdS_5$ solution as $r\to 0$ in the UV
and approach an $AdS_4\times\mathbb{R}$ S-fold solution in the IR, which will be located at $r\to-\infty$. Thus, along the flow we will have $P\ge 0$, with $P$ increasing from zero
as one moves from the UV to the IR.

We now consider the $AdS_4\times\mathbb{R}$ S-fold solutions. In these solutions we have\footnote{Equivalently, we have $C=r/\ell$ after a constant shift of the radial coordinate.} $A= r/\ell$, where $\ell$ is 
a constant and $V$ is constant.
We assume that the $D=5$ dilaton is a linear function of $x$ and all of the scalar fields are constant\footnote{More elaborate S-fold solutions, with the $D=5$ dilaton a linear plus periodic function 
of $x$ are discussed in \cite{Arav:2021tpk}.}. The metric and dilaton of these solutions can thus be written as
\begin{align}\label{fixedptsol}
ds^2&=e^{2V}\left[\ell^2ds^2(AdS_4)-dx^2\right]\,,\nn
\varphi&=k x\,,
\end{align}
where $ds^2(AdS_4)$ has unit radius.
We see that $e^V \ell$ is the radius of the $AdS_4$ with respect to the $D=5$ metric. From \eqref{4dmet} the 
$D=4$ Einstein metric after the Scherk-Schwarz reduction on the $x$ direction is given by 
\begin{align}\label{4dads4}
ds^2_{(4)}=e^{3V}[-dr^2+e^{2r/\ell}e^{-2V}(dt^2 - dy_1^2 - dy_2^2)]\,,
\end{align}
and one sees that $e^{3 V/2}\ell$ is the radius of the $AdS_4$.
 Using the BPS equations, which imply \eqref{beeeq},
its not difficult to show that for such an S-fold solution we have, in particular,
\begin{align}\label{sfoldconds}
k \mu &= \frac{1}{3} e^{\frac{1}{2}\widetilde{\mathcal{K}} +V} = \frac{1}{\ell},\qquad
P= \frac{2}{e^{3 V/2}\ell}\,. 
\end{align}

We now briefly recall  how these $AdS_4\times\mathbb{R}$ S-fold solutions can be used to construct $AdS_4\times S^1\times S^5$ S-fold solutions of type IIB string theory.
One uplifts the $D=5$ solution to obtain an $AdS_4\times\mathbb{R}\times S^5$ solution of type IIB supergravity. One then generates a larger family of type
IIB supergravity solutions by acting with elements of $SL(2,\mathbb{R})$. Within this larger family one constructs the $AdS_4\times S^1\times S^5$ S-fold solutions 
by periodically identifying along the $\mathbb{R}$ direction up to an element $\mathcal{M}\in SL(2,\mathbb{Z})$ in the hyperbolic conjugacy class. For example, we can take $\mathcal{M}$ to be of the form \eqref{genm}
provided that $k\Delta x=\text{arcosh}\,\frac{n}{2}$, where $\Delta x$ is the period over which we S-fold.
These S-fold solutions of type IIB string theory are dual to $d=3$ SCFTs with, generically, $\mathcal{N}=1$ supersymmetry. A key observable for such SCFTs is $\mathcal{F}_{S^3}$, the free energy of the SCFT when placed on $S^3$.
This observable can be obtained from the $D=4$ Newton's constant and, using\footnote{To do this one should identify
$e^{A^{there}}=e^{V}\ell$ and $k^{there}=k\ell$.} \eqref{sfoldconds},
from eq. (3.29) of \cite{Arav:2021tpk} we obtain the result
\begin{align}\label{freeens}
\mathcal{F}_{S^3}
&={N^2}\text{arccosh}\frac{n}{2}\frac{4}{kL^3 P^2}\,,
\end{align}
where $N$ is the quantised flux of the type IIB five-form through the $S^5$.

In the subsequent sections we will construct $D=5$ RG flow solutions that start off at one $AdS_4\times\mathbb{R}$ S-fold solution
in the UV and end up at another $AdS_4\times\mathbb{R}$ S-fold solution in the IR. The procedure described above can be applied to these RG flows
and one obtains RG flow solutions of type IIB string theory that start off at one $AdS_4\times S^1\times S^5$ S-fold solution
in the UV and end up at another $AdS_4\times S^1\times S^5$ S-fold solution the IR, with the entire flow S-folded by the same element $\mathcal{M}\in SL(2,\mathbb{Z})$.
Thus, the free energies in the UV and the IR will be as in \eqref{freeens}, with the same $n$ and $k$ with the values of $P$ given by the $D=5$ solutions.
We will consider such $D=5$ RG flows with the UV at $r=+\infty$ and the IR at $r\to -\infty$. The monotonicity of the gradient flow equations that we deduced from
\eqref{deepeecond} then immediately implies the expected result that
\begin{align}
\mathcal{F}_{S^3}|_{IR}<\mathcal{F}_{S^3}|_{UV}\,.
\end{align}

We will also construct RG flows that start off at the $AdS_5$ vacuum in the UV at $r=0$ and flow to $AdS_4\times\mathbb{R}$ S-fold solutions in the IR at $r=-\infty$.
In an analogous manner the S-folding procedure can also be applied to the uplifted solutions and we obtain RG flows that flow from
$AdS_5\times S^5$ in the UV, suitably deformed and S-folded, ending up at an $AdS_4\times S^1\times S^5$ S-fold solution in the IR. Once again the entire 
flow is S-folded by the same element $\mathcal{M}\in SL(2,\mathbb{Z})$.

\section{Simple $AdS_5$ to S-fold RG flows}\label{lin_dil_flow}

We now begin our construction of BPS RG flow solutions that start off at $AdS_5$ in the UV, deformed
by a linear dilaton source, which then flow to $\mathcal{N}=4,2,1$ $AdS_4\times\mathbb{R}$ S-fold solutions
with a linear dilaton in the IR. The simplest models to construct such solutions are the truncations
denoted by the blue boxes in figure \ref{truncdiag} and this is precisely what will be considered in this section. 

We begin, however, with some general comments which are also relevant for the more general flows constructed in 
section \ref{moreelab}.
All of the $AdS_4\times\mathbb{R}$ S-folds solutions of interest are 
invariant under the discrete symmetry \eqref{discbps} of the BPS equations, $z^A \leftrightarrow -\zb^A$.
If we also change the sign of the $x$ coordinate, the combined transformation leaves
$k$ invariant and just changes the signs of the scalars $\alpha_i$ and also $f$ (appearing in \eqref{dilansatz}).
In the remainder of the paper we will only consider RG flow solutions that
are invariant under this transformation and take
\begin{align}\label{simpansrgflow}
\alpha_i=f=0\,.
\end{align}
Note that the constraint \eqref{consteq} is automatically satisfied for such configurations.

In appendix \ref{specrta} we have analysed the linearised modes of the BPS equations about the various S-folds solutions.
This spectrum is essential for understanding the behaviour of RG flows with S-folds appearing either in the IR or the UV.
We next examine the behaviour of the BPS equations \eqref{eq:10scalarGradientFlowEqst} (or equivalently \eqref{radbpseqst}) 
near an $AdS_5$ boundary located at $r \rightarrow 0^-$, with metric given by
\begin{align}\label{ads5metexp}
ds^2&= \frac{L^2}{r^2}(dt^2 - dy_1^2 - dy_2^2 - dx^2 - dr^2)+\dots \,.
\end{align}
After changing coordinates to a proper distance gauge 
with $g_{\bar r \bar r}=1$, via
\begin{align}
-\frac{r}{L} = e^{-\rb/L}  + \Big( \frac{1}{18}(\phi_{1(s)}^2 + \phi_{2(s)}^2 + \phi_{3(s)}^2)- \frac{k^2L^2}{8} \Big)e^{-3\rb/L}+\ldots\,,
\end{align}
with the assumption of \eqref{simpansrgflow} we find the expansion for the remaining 7 scalar fields is given by 
\begin{align}\label{genscexp}
\varphi&=kx\,,\nn
\phi_1 &=  \phi_{1(s)} e^{-\rb/L} - \frac{\rb}{L}e^{-3\rb/L} \frac{4}{3} \left( 2\phi_{1(s)}^2 - \phi_{2(s)}^2 - \phi_{3(s)}^2\right) + \phi_{1(v)} e^{-3\rb/L} +\ldots 
\,,\nonumber \\
\phi_2 &=  \phi_{2(s)} e^{-\rb/L} - \frac{\rb}{L}e^{-3\rb/L} \frac{4}{3} \left( 2\phi_{2(s)}^2 - \phi_{1(s)}^2 - \phi_{3(s)}^2\right) + \phi_{2(v)} e^{-3\rb/L} +\ldots 
\,,\nonumber \\
\phi_3 &=  \phi_{3(s)} e^{-\rb/L} - \frac{\rb}{L}e^{-3\rb/L} \frac{4}{3} \left( 2\phi_{3(s)}^2 - \phi_{1(s)}^2 - \phi_{2(s)}^2\right) + \phi_{3(v)} e^{-3\rb/L} +\ldots 
\,,\nonumber \\
\phi_4  &=  \phi_{4(s)} e^{-\rb/L} - \frac{\rb}{L}e^{-3\rb/L} \frac{4}{3} \left( \phi_{1(s)}^2 + \phi_{2(s)}^2 + \phi_{3(s)}^2 - 3 \phi_{4(s)}^2 \right) + \phi_{4(v)} e^{-3\rb/L} +\ldots 
\,,\nonumber \\
\beta_i &= \beta_{i(s)} \frac{\rb}{L} e^{-2\rho/L} + \beta_{i(v)} e^{-2\rb/L}+\ldots\,.
\qquad i=1,2\,.
\end{align}
The terms with the $(s)$ subscript fix the sources of the dual operators given in \eqref{opfieldmapz} and so we also have, of course, $\varphi_{(s)}=kx$. 
The terms with
the $(v)$ subscript, which we loosely refer to as ``vevs", determine the expectation values of the dual operators when combined with the source 
terms, as explained\footnote{The holographic renormalisation in \cite{Arav:2020obl,Arav:2021tpk} involves a number of finite counter terms which we do not discuss in this paper.}   in detail in \cite{Arav:2020obl,Arav:2021tpk}.
The BPS equations also imply the following BPS relations between the sources, given by
\begin{align}\label{eq:ten_scalar_source}
\phi_{4(s)} = \frac{k L}{2}, \qquad \beta_{1(s)} = \frac{1}{3}({\phi_{1(s)}^2 + \phi_{2(s)}^2 - 2 \phi_{3(s)}^2}), \qquad \beta_{2(s)} = \phi_{1(s)}^2 - \phi_{2(s)}^2,
\end{align}
and the vevs:
\begin{align}\label{eq:ten_scalar_vev}
\phi_{1(v)} &= k L \phi_{2(s)}\phi_{3(s)}-\phi_{1(s)} \Big( \frac{3k^2L^2}{8}  - \frac{1}{6}({ \phi_{3(s)}^2 + \phi_{2(s)}^2 - 7 \phi_{1(s)}^2}) + 2 (\beta_{1(v)} + \beta_{2(v)} )  \Big),
\nonumber \\
\phi_{2(v)} &= k L \phi_{1(s)}\phi_{3(s)}-\phi_{2(s)} \Big( \frac{3k^2L^2}{8} - \frac{1}{6}( \phi_{1(s)}^2 + \phi_{3(s)}^2 - 7 \phi_{2(s)}^2) + 2 (\beta_{1(v)} - \beta_{2(v)} )  \Big),
\nonumber \\
\phi_{3(v)} &= k L \phi_{1(s)}\phi_{2(s)}-\phi_{3(s)} \Big( \frac{3k^2L^2}{8}  - \frac{1}{6}(\phi_{1(s)}^2 + \phi_{2(s)}^2 - 7 \phi_{3(s)}^2)  - 4 \beta_{1(v)} \Big).
\end{align}
Thus, generically, with \eqref{simpansrgflow}, in addition to $\varphi_{(s)}=kx$ we have three independent sources (e.g. $\phi_{1,2,3(s)}$) and three independent vevs (e.g. $\phi_{4(v)},~\beta_{1,2(v)}$).

Using the dictionary below \eqref{opfieldmapz}, these sources can be directly compared with the field theory analysis made in section \ref{sec:2}.
With $\alpha_i=0$ we have $D^{23}{}_{14}=D^{31}{}_{24}=D^{12}{}_{34}=0$ as well as
\begin{align}\label{dictionary}
(m_1,m_2,m_3,m_4)&\leftrightarrow (\phi_{1(s)},\phi_{2(s)},\phi_{3(s)},-\phi_{4(s)})\,,\nn
\begin{pmatrix}D^{12}{}_{12}\\D^{13}{}_{13}\\D^{14}{}_{14}
\end{pmatrix}
&\leftrightarrow
\begin{pmatrix}\phi_{1(s)}^2+\phi_{2(s)}^2-2\phi_{3(s)}^2\\ \phi_{1(s)}^2+\phi_{3(s)}^2-2\phi_{2(s)}^2\\\phi_{2(s)}^2+\phi_{3(s)}^2-2\phi_{1(s)}^2
\end{pmatrix}\,,
\end{align}
where each identification is up to an overall factor. This is in alignment with the $\cN=1$ deformations
\eqref{emmsn1}-\eqref{deesn1} when $\ln g$ is a linear function and the masses are constant. There is an interesting
feature in the identification for the $\cN=4$ case and  $\cN=2$ case, that we highlight in sections \ref{so3so3trunc} and
\ref{u1u1sec}, respectively.

From the various solutions that we construct, we can construct additional solutions using the $\mathbb{Z}_2\times S_4$ symmetry of the 10-scalar model. Recall that to simplify the presentation we have taken $k>0$. Under the $\mathbb{Z}_2$, which takes $z^A\to- z^A$, we should flip the sign of $k$; we shall not explicitly display such solutions in any of our plots. 
In some of the sub-truncations that we consider there is a remaining $\mathbb{Z}_2\subset S_4$ symmetry associated with flipping the
signs of any two of $(\phi_1,\phi_2,\phi_3)$, which leaves the moment map \eqref{mommapcon} invariant, and we will display the solutions obtained by this action in the plots.

Generically, the RG flow solutions preserve $\mathcal{N}=1$ Poincar\'e supersymmetry in $d=3$. As explained in appendix \ref{appc},
exceptions to this involve the $U(1)\times U(1)$ 6-scalar truncation and the $SU(2)\times U(1)$ 4-scalar truncation, for which the RG flows generically preserve $\mathcal{N}=2$, as well as the $SO(3)\times SO(3)$ invariant 3-scalar truncation for which 
$\mathcal{N}=4$ supersymmetry is preserved.

\subsection{$AdS_5$ to $\cN=4$ S-fold: $SO(3)\times SO(3)$ invariant truncation}\label{so3so3trunc}
We first consider the BPS equations in the $SO(3)\times SO(3)$ invariant 3-scalar truncation that contains the known
$\mathcal{N}=4$ S-fold solution \cite{Inverso:2016eet,Assel:2018vtq}. In this case we are able to construct an analytic RG flow.
This truncation is obtained by taking
\begin{align}\label{eq:so3so3_trunc}
z^2=z^3=-z^4,\qquad  z^2=\bar z^2\,,\qquad \beta_i=0\,,
\end{align}
and keeps 3 of the 10 scalar fields:
$\phi_1=\phi_2=\phi_3=-\phi_4$, $\alpha_1=\alpha_2=\alpha_3$ and $\varphi$. 
As we explain in appendix \ref{appc}, solutions to the BPS equations \eqref{eq:10scalarGradientFlowEqst} (or equivalently \eqref{radbpseqst}) for this truncation preserve $\mathcal{N}=4$ Poincar\'e supersymmetry in $d=3$.

We now analyse the BPS equations \eqref{eq:10scalarGradientFlowEqst}, along
with the further ansatz \eqref{simpansrgflow}. From the BPS equation for $z^2$, the reality condition on $z^2$ in \eqref{eq:so3so3_trunc}
leads to an algebraic constraint which may be used to solve for $V$,
\begin{align}
e^{V}=-\sqrt{2} kL \frac{\sqrt{\cos4\phi_1(1+\cos4\phi_1)}}{\sin 4\phi_1}\,.
\end{align}
From the BPS equations \eqref{radbpseqst} for $z^1$ we then obtain
\begin{align}\label{bpsn4flow}
4 \phi_1'-k (3 \cos 4 \phi_1-1)&=0\,.
\end{align}
Having solved this one can obtain $A$ by integrating the first equation in 
 \eqref{radbpseqst}.
 
In fact we can solve \eqref{bpsn4flow} analytically and we find
\begin{align}\label{phi1sol}
\phi_1=\frac{1}{2} \tan ^{-1}\Big(\frac{\tanh \left(\sqrt{2} k r\right)}{\sqrt{2}}\Big)\,,
\end{align}
where we have dropped a constant by shifting $r$. The 
metric functions which complete the solution are then given by
\begin{align}
e^{2A}&=\frac{2k^2L^2}{\sinh^2(\sqrt{2} kr)}\,,\qquad
e^{2V}=L^2k^2(1+\frac{2}{\sinh^2(\sqrt{2} kr)})\,,
\end{align}
where we have dropped a trivial integration constant in $A$ which can be absorbed by a rescaling of 
the coordinates $(t,y_1,y_2)$ in \eqref{iso21metans}.

This exact solution describes an RG flow from $AdS_5$ in the UV, located at $r=0$, down to
the $\mathcal{N}=4$ $AdS_4\times\mathbb{R}$ S-fold solution in the IR at $r\to-\infty$.
 Indeed as 
$r\to -\infty$ the solution asymptotes to
\begin{align}\label{nfoursfoldsol}
ds^2&\to k^2 L^2[8e^{2\sqrt{2} k r}(dt^2 - dy_1^2 - dy_2^2)-dr^2 -dx^2]+\dots \,,\nn
\phi_1&\to-\frac{1}{2}\cot^{-1}\sqrt{2}+\frac{\sqrt 2}{3}e^{2\sqrt{2}kr}+\dots \,,\qquad
\varphi=k x\,.
\end{align}
This is of the form \eqref{fixedptsol} with $e^V= k L$, $\ell=1/(\sqrt{2}k)$
and we immediately see that we are indeed approaching the
$\mathcal{N}=4$ $AdS_4\times \mathbb{R}$ S-fold solution of \cite{Inverso:2016eet} as written
in \cite{Arav:2021tpk}. In particular the free energy is given
by \eqref{freeens} with 
$4/({kL^3 P^2})=1/2$
in agreement with \cite{Assel:2018vtq}. 
Recalling \eqref{4dads4}, with respect to the $D=4$ Einstein metric the $AdS_4$ radius is
$e^{3 V/2}\ell$, the sub-leading behaviour of the scalar field is associated with an irrelevant operator in the
$d=3$ SCFT with scaling dimension $\Delta=5$ (see appendix \ref{specrta}).

We now examine the behaviour of the RG flow solution as we approach the UV.
As $r\to 0^-$ it approaches $AdS_5$ as in \eqref{ads5metexp}, \eqref{genscexp}
along with the relations \eqref{eq:ten_scalar_source}, \eqref{eq:ten_scalar_vev}.
The RG flow is uniquely fixed by the $D=5$ dilaton source, $\varphi=kx$ and we have
the following non-vanishing sources
and vevs
 for the remaining scalars
\begin{align}\label{n4scesvevs}
\phi_{1(s)} &=\phi_{2(s)} =\phi_{3(s)} =-\phi_{4(s)} =- \frac{kL}{2},\nn
\phi_{1(v)} &=\phi_{2(v)} = \phi_{3(v)} = -\phi_{4(v)} =  \frac{13}{24}k^3L^3.
\end{align}

Recalling \eqref{dictionary}, we see these source terms are in alignment with our analysis
of supersymmetric deformations of $\mathcal{N}=4$ SYM theory in section \ref{sec:2}; see
\eqref{emmsn4}-\eqref{deesn4}. Interestingly, this solution satisfies the condition \eqref{n4cond} for the signs of the masses which, from
the boundary point of view, is not required when $g''=0$ (associated with a linear dilaton in the bulk solution).  
We have also constructed an analytic solution\footnote{
This solution, which exists in a \emph{different} $\cN=4$ truncation given by $z^1=-\zb^2=-\zb^3=+\zb^4,~\beta_i=0 \Leftrightarrow \alpha_i = \beta_i=0,~\phi_1=\phi_2=\phi_3=+\phi_4$, is  $\phi_1= - \frac{ k r}{2},~ e^{2A} = \frac{k^2 L^2}{\tan(kr)},~ e^{2V} = \frac{k^2L^2}{\sin(kr)}$.} that is associated with sources 
$\phi_{1(s)} =\phi_{2(s)} =\phi_{3(s)} =+\phi_{4(s)} =- \frac{kL}{2}$, and hence associated with violating the condition \eqref{n4cond}, but we find that it flows to a naked singularity in the IR. Moreover, there is no S-fold solution with 
$\phi_{1} =\phi_{2} =\phi_{3} =+\phi_{4}$.
It would be interesting to understand the physical reason underlying this differing behaviour.

\subsection{$AdS_5$ to $\cN=2$ S-fold: $SU(2)\times U(1)$ invariant truncation}\label{su2u1}

We next consider the BPS equations in the $SU(2)\times U(1)$ invariant 4-scalar truncation which contains the known
$\mathcal{N}=2$ S-fold solution \cite{Guarino:2020gfe,Bobev:2020fon}. This truncation is obtained by taking
\begin{align}\label{eq:su2u1_trunc}
z^2=-z^4,\qquad  z^1=-z^3,\qquad z^2=\bar z^2\,, \qquad \beta_2=0
\end{align}
and keeps 4 of the 10 scalar fields:
$\phi_3=-\phi_4$, $\alpha_3$, $\varphi$ and $\beta_1$.
As we explain in appendix \ref{appc}, solutions to the BPS equations \eqref{eq:10scalarGradientFlowEqst} 
for this truncation preserve $\mathcal{N}=2$ Poincar\'e supersymmetry in $d=3$.

We now examine the BPS gradient flow equations \eqref{eq:10scalarGradientFlowEqst}  with the
further ansatz \eqref{simpansrgflow}. The reality condition for $z^2$ implies the constraint
\begin{align}
e^V=-\frac{kLe^{4\beta_1}}{\tan 2\phi_3}\,.
\end{align}
After eliminating $V$ from the remaining equations we are left with solving
\begin{align}\label{bpsn2}
\phi_3' &=  k(e^{6\beta_1}\cos 2\phi_3 - \frac{1}{2})\,,\nn
\beta_1' &= k \frac{e^{6\beta_1} - \cos 2\phi_3}{3\sin 2\phi_3}, \nn
 C' &= -k\frac{2e^{6\beta_1} +\cos 4\phi_3 \sec 2 \phi_3}{2 \sin 2\phi_3}.
\end{align}

This system has a unique $AdS_4\times \R $ S-fold (recall we have assumed positive $k$)
with $C=r/\ell$ and constant scalars, given by
\begin{align}\label{n2sfoldsol}
ds^2&\to \frac{L^2k^2}{2^{2/3}}[\frac{1}{k^2}ds^2(AdS_4)-dx^2]\,,\nn
\phi_3 &= -\frac{\pi}{8}\,,\qquad  e^{6\beta_1} = \frac{1}{\sqrt 2}\,,\qquad
\varphi=k x\,.
\end{align}
This is of the form \eqref{fixedptsol} with $e^{V}=Lk/2^{1/3}$, $\ell=1/k$ and 
 we have the $\mathcal{N}=2$ $SU(2)\times U(1)$ invariant S-fold solution of \cite{Guarino:2020gfe,Bobev:2020fon}
 as written in \cite{Arav:2021tpk}.
The free energy is given by
\eqref{freeens} with 
$4/({kL^3 P^2})=1/2$.

\begin{figure}[h!]
\centering
{
\includegraphics[scale=0.55]{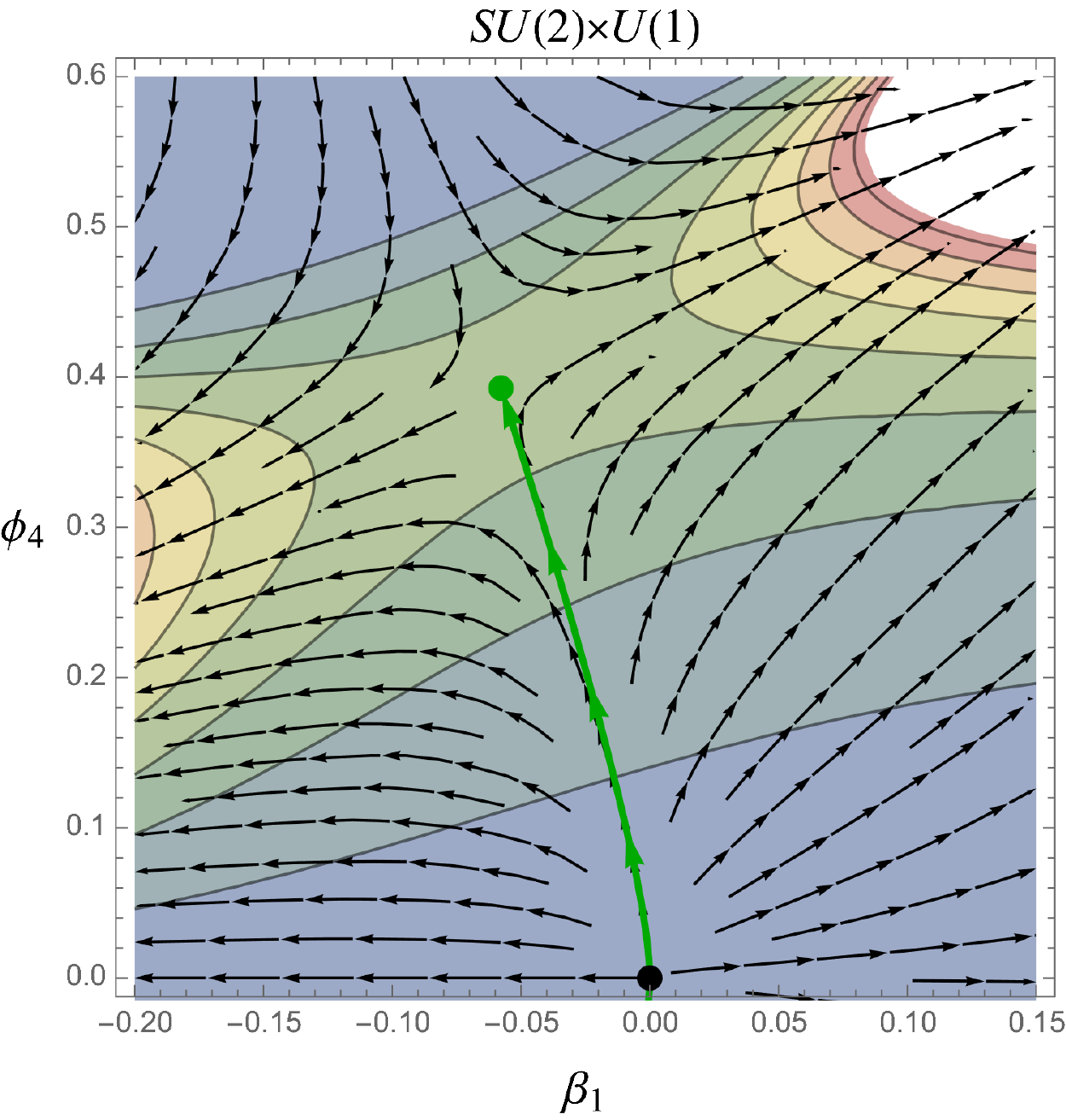}\qquad
\includegraphics[scale=0.55]{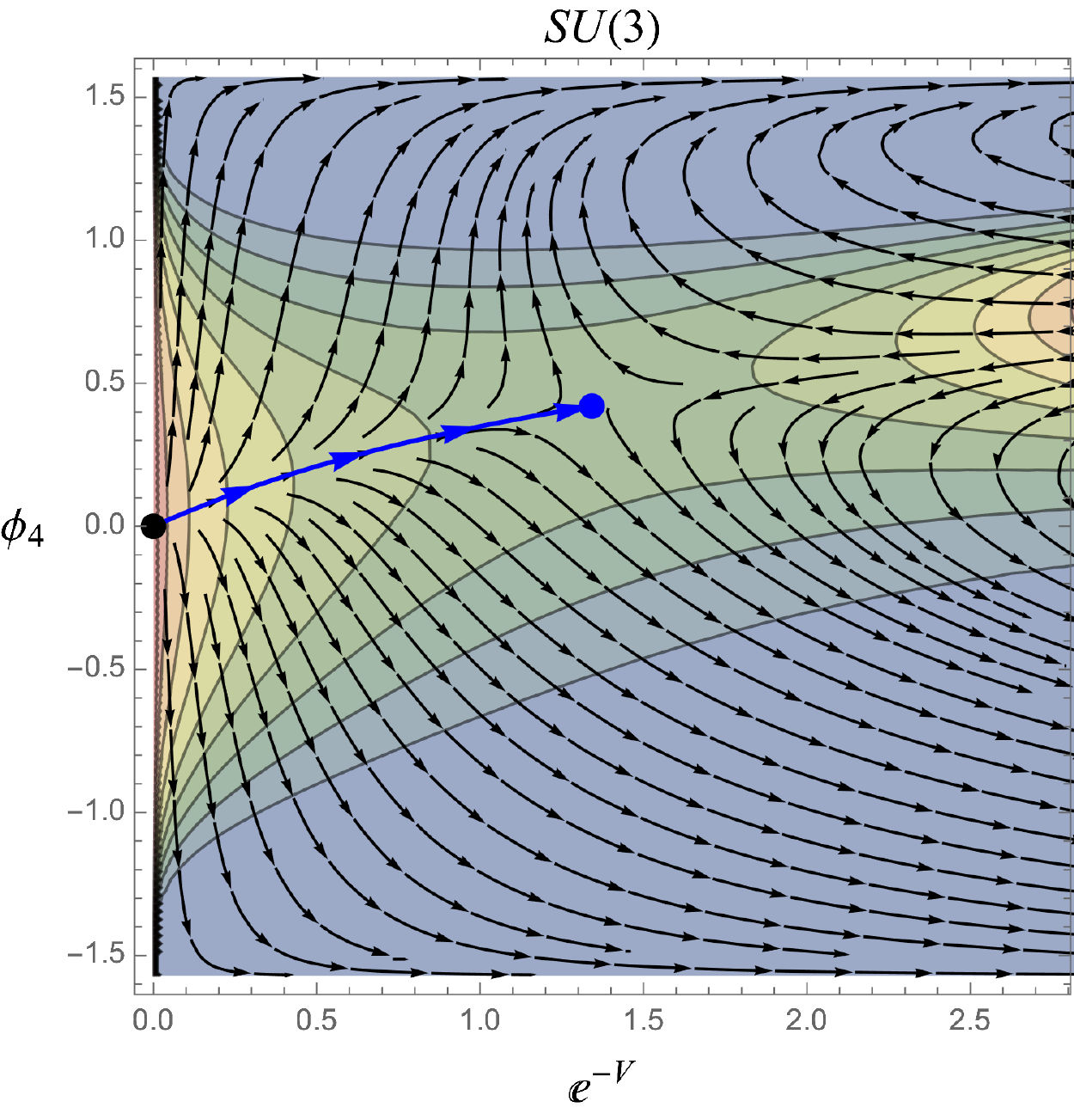}
}
\caption{RG flows from $AdS_5$ to $AdS_4\times\mathbb{R}$ S-folds. The stream plots are in
the direction of decreasing $r$ and the contours are constant lines of $P$. The left plot is for the $SU(2)\times U(1)$ invariant truncation: 
the green dot is the $SU(2)\times U(1)$ invariant $\mathcal{N}=2$ S-fold, the black dot is the $AdS_5$ vacuum and one sees a unique RG flow connecting them.
The right plot is for the $SU(3)$ invariant truncation: the blue dot is the $\mathcal{N}=1$ S-fold and one sees a unique RG flow
connecting it with the $AdS_5$ vacuum solution, again marked with a black dot, which is located at $V\to \infty$.}
\label{4scalar}
\end{figure}

We next examine the behaviour of the BPS equations as we approach the $AdS_5$ boundary in the UV
as $r\to 0^-$. From \eqref{eq:ten_scalar_source}, \eqref{eq:ten_scalar_vev} we 
deduce the following relations between the sources and the vevs for the scalars
\begin{align}\label{scesn2}
\phi_{3(s)} &= -\phi_{4(s)} =-\frac{kL}{2}, \qquad \beta_{1(s)}= - \frac{k^2 L^2}{6}, \nn 
\phi_{3(v)} &= -\phi_{4(v)} = \frac{k^3 L^3}{3} - 2 k L \beta_{1(v)}.
\end{align}
Thus, $\varphi=kx$ is the only independent source and there is one independent vev (e.g. $\beta_{1(v)}$) which then determines
the RG flow from the $AdS_5$ boundary. 
This is consistent with the fact that we have one physical integration constant for the system \eqref{bpsn2}, after taking into account
that we can shift the radial coordinate and absorb a constant shift of $A$ by rescaling $(t,y_1,y_2)$ in \eqref{iso21metans}. 
 In figure \ref{4scalar} we display a stream plot associated with the first two BPS equations in \eqref{bpsn2}. This clearly shows that
 there is a unique flow\footnote{The numerical value of $\beta_{1(v)}$ can be determined by a more careful numerical analysis which we have not carried out.}
 from the $AdS_5$ boundary to the $\mathcal{N}=2$ $AdS_4\times \R $ S-fold. 
In constructing this solution numerically we utilised the fact that 
in the linearised fluctuations about the $\mathcal{N}=2$ S-fold solution in the IR there is a single mode, 
associated with an irrelevant IR operator with $d=3$ scaling dimension
$\Delta = \frac{5+\sqrt{17}}{2}$ (see appendix \ref{specrta}), which can be used to flow from the IR out to the $AdS_5$ in the UV.
Finally, recalling \eqref{dictionary} we see that the sources in \eqref{scesn2} are in alignment with
\eqref{emmsn2}-\eqref{ntwocond}, with $m_1=m_2=0$.

\subsection{$AdS_5$ to $\cN=1$ S-fold: $SU(3)$ invariant truncation}\label{su3}
We now consider the BPS equations in the $SU(3)$ invariant 2-scalar truncation that contains the known
$\mathcal{N}=1$ $SU(3)$ invariant S-fold solution \cite{Guarino:2019oct}. This truncation is obtained by taking
\begin{align}\label{eq:su3_trunc}
z^1=-z^2=-z^3=z^4,\qquad \beta_i=0\,,
\end{align}
and keeps 2 of the 10 scalar fields: $\phi_4$, and $\varphi$.
Solutions to the BPS equations \eqref{eq:10scalarGradientFlowEqst} for this truncation generically preserve $\mathcal{N}=1$ Poincar\'e supersymmetry in $d=3$.

In this model there is no reality constraint on the $z^A$ fields and so we cannot solve for $V$ algebraically.
Instead we find the BPS gradient flow equations \eqref{eq:10scalarGradientFlowEqst}  are given by
\begin{align}
 \phi_4' = & -\frac{3 e^{V} \tan \phi_4}{L} + k ,\nonumber \\
 V' = & \frac{e^{V}}{\cos^2\phi_4 L} - 2 k  \tan \phi_4 , \nonumber\\
C' = & \frac{3 e^{V}}{2\cos^2 \phi_4 L} - k  \tan \phi_4.
\end{align}

This system has a unique $AdS_4\times \R $ S-fold (recall we have assumed positive $k$)
with $C=r/\ell$ and constant scalars, given by
\begin{align}\label{n1sfoldsol}
ds^2&\to \frac{5L^2k^2}{9}[\frac{5}{4k^2}ds^2(AdS_4)-dx^2]\,,\nn
\phi_4 &= \cos^{-1}\sqrt{5/6}\,,\qquad
\varphi=k x\,.
\end{align}
This of the form \eqref{fixedptsol} with $e^{V}=(\sqrt{5}/3)Lk$, $\ell=\sqrt{5}/(2k)$ 
and  we have the $\mathcal{N}=1$ $SU(3)$ invariant S-fold solution of \cite{Guarino:2019oct}
as written in \cite{Arav:2021tpk}.
The free energy is given by
\eqref{freeens} with 
$4/({kL^3 P^2})=5^{5/2}/108$.

We now examine the behaviour of the BPS equations as we approach the $AdS_5$ boundary in UV
as $r\to 0^-$. From \eqref{eq:ten_scalar_source}, \eqref{eq:ten_scalar_vev} we deduce that the 
RG flow is uniquely fixed by the $D=5$ dilaton source, $\varphi=kx$ with the remaining source fixed to be
\begin{align}\label{simpsce}
\phi_{4(s)} = \frac{kL}{2}\,.
\end{align}
There is only one non-vanishing vev $\phi_{4(v)}$ which determines
the flows from the $AdS_5$ boundary.
This is consistent with the fact that we again have one physical integration constant for the system \eqref{bpsn2}, after taking into account
that we can shift the radial coordinate and absorb a constant shift of $A$ by rescaling $(t,y_1,y_2)$ in \eqref{iso21metans}. 
 In figure \ref{4scalar} we display a stream plot associated with the first two BPS equations in \eqref{bpsn2}. This clearly shows that
 there is a unique flow from the $AdS_5$ boundary to the $\mathcal{N}=1$ $AdS_4\times \R $ S-fold. To construct this solution numerically
 we can utilise the fact that 
in the linearised fluctuations about the S-fold there is a single mode, 
associated with an irrelevant IR operator with $d=3$ scaling dimension
$\Delta= 2+\sqrt{6}$ (see appendix \ref{specrta}), which can be used to shoot out from the IR to hit $AdS_5$ in the UV.
Finally, recalling \eqref{dictionary} we see that the sources in \eqref{simpsce} are in alignment with
\eqref{emmsn1}-\eqref{deesn1}, with $m_1=m_2=m_3=0$.

\section{Marginal deformations of S-folds and more elaborate RG flows}\label{sec:more elaborate}
In this section we construct various solutions to the BPS equations \eqref{eq:10scalarGradientFlowEqst}
for the consistent truncations denoted by the red boxes in figure \ref{truncdiag}. 

\subsection{Marginal deformations and RG flows in the $U(1)\times U(1)$ invariant truncation}\label{u1u1sec}

The  $U(1)\times U(1)$ invariant 6-scalar truncation is obtained by taking
\begin{align}\label{eq:u1u1_trunc}
z^2=\zb^2 = -z^4  = -\zb^4, \qquad \beta_2 = 0\,,
\end{align}
and keeps 6 of the 10 scalar fields: $\phi_1 = \phi_2$ , $\phi_3 = -\phi_4$, $\alpha_1=\alpha_2$, $\alpha_3$, $\beta_1$ and $\varphi$.
In addition to the $\mathbb{Z}_2$ symmetry given by $z^A\to -z^A$, this model has another $\mathbb{Z}_2$ symmetry, arising from the $S_4$ discrete symmetries of the 10-scalar model, which flips the sign of $\phi_1$ and $\phi_2$.
As we explain in appendix \ref{appc}, solutions to the BPS equations \eqref{eq:10scalarGradientFlowEqst} (or equivalently \eqref{radbpseqst}) for this truncation preserve $\mathcal{N}=2$ Poincar\'e supersymmetry in $d=3$.
We continue to impose $\alpha_i=f=0$. 

\begin{figure}[h!]
\centering
{\includegraphics[scale=1.2]{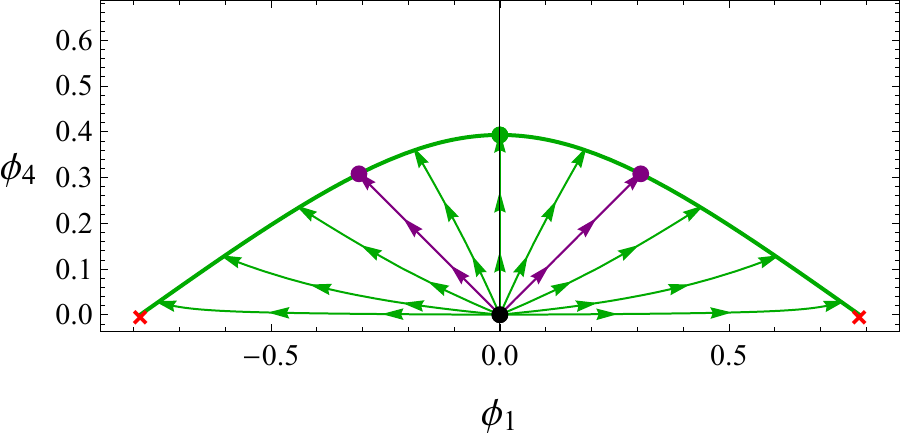}}
\caption{Conformal manifold and parametric plots for the RG flows in the $U(1)\times U(1)$ invariant 6-scalar truncation. The black dot is the $AdS_5$ vacuum, 
the green dot is the $\mathcal{N}=2$, $SU(2)\times U(1)$ invariant $AdS_4\times \mathbb{R}$ S-fold solution and the two purple dots are
$\mathcal{N}=4$, $SO(3)\times SO(3)$ invariant $AdS_4\times \mathbb{R}$ S-fold solutions. The solid green line is the new one parameter family of $\mathcal{N}=2$ S-fold solutions and the red crosses represent singular solutions. The vertical RG flow to the 
$\mathcal{N}=2$ S-fold was considered in section \ref{su2u1}, while the left purple line is the RG flow to the 
$\mathcal{N}=4$ S-fold discussed in section \ref{so3so3trunc}.}\label{u1u1flows}
\end{figure}

A striking feature of this model is that it contains a new one parameter family of 
S-fold solutions, which generically preserve $\mathcal{N}=2$ supersymmetry.
Specifically, we have
\begin{align}\label{confman}
A &= r/\ell, \qquad k \ell = \tan 2 \phi_4, \qquad e^{2V} = \frac{L^2 \tan 2 \phi_4^{2/3}}{2^{2/3} \ell^2}, \nn
e^{6\beta_1} &= \frac{\cos 2 \phi_4}{\cos 2 \phi_1}, \qquad  \cos 4 \phi_1 + 2 \cos 4 \phi_4 = 1\,,\qquad \varphi=kx\,,
\end{align}
along with $\alpha_i=0$.
This family contains the previously identified $\mathcal{N}=2$, $SU(2) \times U(1)$ invariant S-fold at $\phi_1=0,~\phi_4 =  \pi/8$, as well as the 
$\mathcal{N}=4$, $SO(3)\times SO(3)$ invariant S-fold at $\pm \phi_1= \phi_4 = \text{arccos}(1/3)/4$.
When restricting to $k>0$ we take $0<\phi_4 \le \pi/8$, with $\phi_1\in(-\pi/4,\pi/4)$ and we notice that as $\phi_4  \rightarrow 0$ the geometry becomes singular since $e^V \rightarrow 0,~  k \ell \rightarrow 0$. This family of solutions is denoted by the solid green line in figure \ref{u1u1flows}.

This one parameter family of S-folds is dual to a subset of the conformal manifold that contains\footnote{The family is clearly different to the one parameter family of $\cN=2$ S-folds constructed in \cite{Guarino:2020gfe}, which does not contain the $\cN=4$ S-fold.}  
the $\cN=4$ S-fold. Since conformal manifolds for $d=3$ SCFTs that preserve $\mathcal{N}=2$ supersymmetry are known to be 
K\"ahler, it must be possible to find at least one other exactly marginal deformation. To find it we need to enlarge the consistent truncation that
we have been considering to include extra fields. It seems likely that there is an extra axionic type field that can be used to generate the extra deformations and would provide an additional compact direction. Since our one-parameter family of solutions ends at singular configurations, 
it seems plausible that the K\"ahler space has some kind of cusp-like singularity.

The free energy for the whole family of S-fold solutions is exactly the same and
given by \eqref{freeens} with $4/({kL^3 P^2})=1/2$. The fact that 
the free energy for the $\mathcal{N}=2$ $SU(2)\times U(1)$ invariant S-fold solution of \cite{Guarino:2020gfe,Bobev:2020fon} is the same as that of the $\mathcal{N}=4$ S-fold was highlighted in \cite{Bobev:2020fon}. Two possibilities for why this could be the case were discussed in \cite{Bobev:2020fon}, including the possibility that the two S-folds lie on the same conformal manifold, as we have explicitly demonstrated here.

We now construct a one parameter family of RG flow solutions that flow between the vacuum $AdS_5$ in the UV to each of these
$AdS_4\times\mathbb{R}$ S-fold solutions in the IR.
With $\alpha_i=f=0$, the first order BPS equations \eqref{eq:10scalarGradientFlowEqst}
involve 5 real functions: $A,V, \phi_1,\phi_4,\beta_1$, along with $\varphi=k x$. As there is a reality constraint on the $z^A$ fields for this truncation we can solve for $V$ algebraically. Taking into account that we can shift the radial coordinate and absorb a constant shift of $A$ by rescaling $(t,y_1,y_2)$ in \eqref{iso21metans}, we deduce that a solution will be specified by specifying two integration constants.
The expansion about the $AdS_5$ vacuum in the UV is given by
\eqref{genscexp}-\eqref{eq:ten_scalar_vev} with 
\begin{align}\label{scesu1u1gen}
\phi_{4(s)} &= \frac{kL}{2}, \qquad \beta_{1(s)} = \frac{2}{3}\phi_{1(s)}^2 - \frac{k^2L^2}{6}\,, \nn
\phi_{1(v)} &= - \phi_{1(s)} \Big(   \phi_{1(s)}^2 + \frac{5}{6} k^2 L^2 +2 \beta_{1(v)}   \Big), \quad
\phi_{4(v)} = - k L \Big(  \frac{k^2L^2}{3}+ \frac{5}{6}{\phi_{1(s)}^2}-2 \beta_{1(v)}    \Big)
\end{align}
Thus in addition to the $D=5$ dilaton source, $\varphi=kx$ there is one other independent source $\phi_{1(s)}$,
and there is also one independent vev $\beta_{1(v)}$, and these are the two integration constants that specify an RG flow solution.

In fact for a given $\phi_{1(s)}/k$, we find that there is just one RG flow solution that approaches an S-fold solution. Furthermore,
as we vary $\phi_{1(s)}/k$ we  cover the entire conformal manifold. As discussed in appendix \ref{specrta}
within the $U(1)\times U(1)$ invariant truncation each $\cN=2$ S-fold solution has 
an irrelevant BPS source which we can turn on, with 
\begin{align}
\Delta= \frac{5}{2} + \sqrt{\frac{49}{4}+\frac{16}{\cos 4\phi_1-3}} > 3\,.
\end{align}
Numerically we can construct an RG flow to the $AdS_5$ vacuum in the UV by turning on this irrelevant source in the IR. Parametric plots for these RG flows are presented in figure \ref{u1u1flows}.
Finally, recalling \eqref{dictionary} we see that the sources in \eqref{scesu1u1gen} are in alignment with
\eqref{emmsn2}-\eqref{ntwocond}, with $m_1=m_2$ and $\kappa_3=\kappa_4=+1$. In particular, it is interesting to observe that
the condition \eqref{ntwocond} is satisfied which, from the boundary analysis, is required only when $m_1',m_2'\ne 0$.
It would be interesting to have a better understanding of why this is the case.

\subsection{RG flows in the $SU(2)$ invariant truncation}\label{moreelab}
This model contains the $AdS_5$ vacuum, dual to $\mathcal{N}=4$ SYM theory, as well another $AdS_5$ solution
dual to the LS $\mathcal{N}=1$ SCFT. In addition it also contains both of the known $\mathcal{N}=2$, $SU(2)\times U(1)$ invariant and the
$\mathcal{N}=1$, $SU(3)$ invariant $AdS_4\times \mathbb{R}$ S-fold solutions that have linear $D=5$ dilatons.
It is well known that there is an RG flow from the $AdS_5$ vacuum to the LS $AdS_5$ solution \cite{Freedman:1999gp}. In the last section we showed that there are RG flows from the $AdS_5$ vacuum to both of the above $\mathcal{N}=2$ and $\mathcal{N}=1$ S-fold solutions. Thus, we can immediately anticipate that this model will possess an elaborate set of RG flows which, moreover, can exhibit intermediate scaling behaviour
as various fixed point solutions are approached. We now present some details.

The $SU(2)$ invariant 5-scalar truncation is obtained by taking
\begin{align}\label{eq:su2_trunc}
z_3=-z_1,~z_4=-z_2,\qquad \beta_2=0\,,
\end{align}
and keeps 5 of the 10 scalar fields: $\phi_3,\phi_4,\alpha_3,\varphi,\beta_1$.
In addition to the $\mathbb{Z}_2$ symmetry given by $z^A\to -z^A$, this model has another $\mathbb{Z}_2$ symmetry, arising from
the $S_4$ discrete symmetries of the 10-scalar model, which flips the sign of $\phi_3$.
We are interested in solving the BPS equation given in \eqref{eq:10scalarGradientFlowEqst} and we will continue to impose
$\alpha_i=f=0$. 

The $AdS_5$ vacuum solution has vanishing scalar fields and radius $L$. In addition, the model contains two LS $AdS_5$ solutions with
\begin{align}
\phi_3=\pm\frac{\pi}{6},\qquad\beta_1=-\frac{1}{6}\ln 2\,,\qquad \phi_4=\alpha_3=0\,,\qquad \tilde L=\frac{3}{2^{5/3}}L\,,
\end{align}
and constant $\varphi$, 
where $\tilde L$ is the radius of the the $AdS_5$ spacetime.
The model also contains the $\mathcal{N}=2$, $SU(2)\times U(1)$ invariant $AdS_4\times\mathbb{R}$ S-fold solution given in \eqref{n2sfoldsol} (with $\phi_4=-\phi_3$) along with the physically equivalent $\mathcal{N}=2$ S-fold obtained by flipping the sign of $\phi_3$.
In addition, it contains the $\mathcal{N}=1$, $SU(3)$ invariant $AdS_4\times\mathbb{R}$ S-fold given in \eqref{n1sfoldsol} (with $\phi_3=0$).

We now construct a one parameter family of RG flow solutions that flow between the vacuum $AdS_5$ in the UV and one of the 
$\mathcal{N}=2$ $AdS_4\times\mathbb{R}$ S-fold solutions in the IR.
With $\alpha_i=f=0$, the first order BPS equations \eqref{eq:10scalarGradientFlowEqst}
involve 5 real functions: $A,V \phi_3,\phi_4,\beta_1$, along with $\varphi=k x$. As there is no reality constraint on the $z^A$ fields for this truncation we cannot solve for $V$ algebraically. Taking into account that we can shift the radial coordinate and absorb a constant shift of $A$ by rescaling $(t,y_1,y_2)$ in \eqref{iso21metans}, we deduce that a solution will be specified by three integration constants.
The expansion about the $AdS_5$ vacuum in the UV is given by
\eqref{genscexp}-\eqref{eq:ten_scalar_vev} with 
\begin{align}\label{scevevgenn1}
\phi_{4(s)} = \frac{kL}{2}, \quad \beta_{1(s)} =- \frac{2}{3} \phi_{3(s)}^2, \quad \phi_{3(v)} = -\phi_{3(s)} \Big( \frac{7}{6} \phi_{3(s)}^2 +\frac{3k^2L^2}{8}- 4 \beta_{1(v)}\Big)\,.
\end{align}
Thus, in addition to the $\varphi=k x$ deformation, there is one independent source (e.g. $\phi_{3(s)}$) and 
two independent vevs (e.g. $\beta_{1(v)},~\phi_{4(v)}$) and these are the three integration constants that specify an RG flow solution.
In fact for a given $\phi_{3(s)}/k$, there is just one RG flow solution that approaches the $\mathcal{N}=2$ $AdS_4\times\mathbb{R}$ S-fold solution in the IR. One way to see this is to observe that if we expand about the $\mathcal{N}=2$ $AdS_4\times\mathbb{R}$ S-fold 
in the IR there are two irrelevant BPS modes, with $d=3$ conformal scaling dimensions 
$\Delta = \frac{5+\sqrt{17}}{2}$, $\Delta = \frac{3+\sqrt{17}}{2}$ (see appendix \ref{specrta}), which we can use to shoot out from the IR.
A scale invariant ratio of these IR deformations then
parametrises the family of RG flows that hit $AdS_5$ in the UV.
Also, recalling \eqref{dictionary} we see that the sources in \eqref{scevevgenn1} are in alignment with
\eqref{emmsn1}-\eqref{deesn1}, with $m_1=m_2=0$.

\begin{figure}[h!]
\centering
{\includegraphics[scale=0.3]{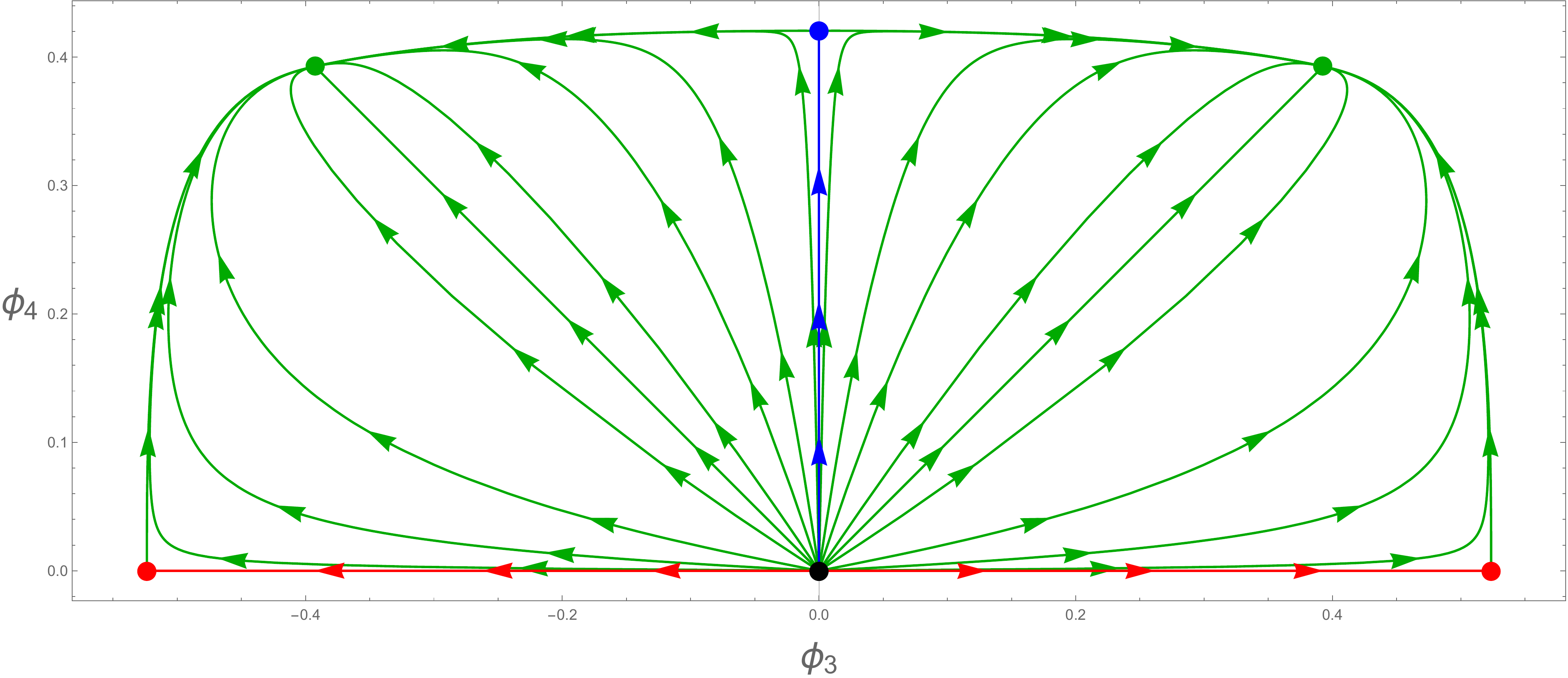}}
\caption{Parametric plots for the RG flows in the $SU(2)$ invariant 5-scalar truncation. The black dot is the $AdS_5$ vacuum, the red dots are two LS $AdS_5$ solutions, the green dots are two $\mathcal{N}=2$, $SU(2)\times U(1)$ invariant $AdS_4\times \mathbb{R}$ S-fold solutions, while the blue dot is the $\mathcal{N}=1$, $SU(3)$ invariant $AdS_4\times S^1$ S-fold solution. The generic RG flows, denoted by green lines, move from the $AdS_5$ vacuum in the UV to the 
$\mathcal{N}=2$ S-fold solution in the IR and there are a number of interesting limiting solutions. 
The green line with $\phi_3=-\phi_4$ is the 
$\mathcal{N}=2$ RG flow considered in section \ref{su2u1}, the blue line with $\phi_3=0$
is the $\mathcal{N}=1$ RG flow considered in section \ref{su3} and the red lines are the homogeneous
RG flows from the $AdS_5$ vacuum to the LS fixed points.
}\label{su2u1flows}
\end{figure}

We have summarised these RG flows as parametric plots in the $\phi_3,\phi_4$ plane, denoted by the green curves in figure \ref{su2u1flows}. 
Associated with the $\mathbb{Z}_2$ symmetry of this model, the figure is symmetric under $\phi_3\to-\phi_3$.
Notice that the diagonal line with $\phi_3=-\phi_4$ is the RG flow
solution to the $\mathcal{N}=2$ S-fold solution in the IR that lies
within the $SU(2)\times U(1)$ invariant truncation that we already discussed in section \ref{su2u1}. 
The line with $\phi_3=+\phi_4$ is a physically equivalent RG flow.
Similarly, the vertical blue line with 
$\phi_3=0$ is the RG flow that goes from the $AdS_5$ vacuum to the $\mathcal{N}=1$ S-fold solution in the IR that lie within the $SU(3)$
invariant truncation that we presented in section \ref{su3}. The horizontal red line is the homogeneous RG flow solution of \cite{Freedman:1999gp}, with vanishing linear dilaton source 
i.e. $k=0$, that approaches the LS $AdS_5$ fixed point in the IR.

The green lines in figure \ref{su2u1flows} reveal a one parameter family of RG flows from $AdS_5$ in the UV to the
$\mathcal{N}=2$ $AdS_4\times\mathbb{R}$ S-fold solution in the IR. The solutions can be parametrised by\footnote{The two integration constants in the IR associated with the two irrelevant operators can be combined to give a dimensionless ratio which can also be used to parametrise the one parameter family of RG flows.} the scale invariant quantity
$ \phi_{3(s)}/k$. The diagonal flow has $\phi_{3(s)}/k=\pm L/2$. As $\phi_{3(s)}/k\to 0$ the solution closely the approaches 
$\mathcal{N}=1$ $AdS_4\times\mathbb{R}$ S-fold solution before heading off to the $\mathcal{N}=2$ $AdS_4\times\mathbb{R}$ S-fold solution
in the far IR. Thus, these limiting RG flows will exhibit an intermediate scaling behaviour that is determined by the
$\mathcal{N}=1$ S-fold solution. 
Furthermore, these limiting RG flows reveal the existence of new RG flows that start from an
$\mathcal{N}=1$ $AdS_4\times\mathbb{R}$ S-fold solution in the UV and flow to an
$\mathcal{N}=2$ $AdS_4\times\mathbb{R}$ S-fold solution in the IR. To construct these
limiting solutions directly, one needs to deform the $\mathcal{N}=1$ $AdS_4\times\mathbb{R}$ S-fold solution in the UV using
the relevant operator with $\Delta=5/3_+$ in table \ref{bpsn1} of appendix \ref{specrta},
and suitably tuning the vevs for the $\Delta=2$ and $\Delta= (1+\sqrt{6})$ modes.

We can also consider the one parameter family of RG flows in figure \ref{su2u1flows} in the limit that 
$\phi_{3(s)}/k\to \infty$. One way this can be achieved is by fixing $\phi_{3(s)}$ and taking $k\to 0$ and so it is perhaps not
too surprising that we then approach the LS $AdS_5$ fixed point before heading off to the $\mathcal{N}=2$ $AdS_4\times\mathbb{R}$ S-fold solution
in the far IR. Associated with this feature, these limiting RG flow solutions will exhibit intermediate scaling that are governed by the
LS fixed point. In addition, we can conclude that there are RG flows that start off at one of the LS fixed points and then flow to 
an $\mathcal{N}=2$ $AdS_4\times\mathbb{R}$ S-fold solution in the IR. To construct these solutions directly,
one needs to deform the  LS fixed point in the UV with  $\varphi=kx$ along with a relevant operator in the LS SCFT with $d=4$ scaling dimension $\Delta=1+\sqrt{7}$, suitably tuning a vev for a $\Delta=3$ operator.

\subsection{RG flows in the $SO(3)$ invariant truncation}

The $SO(3)$ invariant 4-scalar truncation is obtained by taking
\begin{align}\label{eq:so3_trunc}
z_2 = z_3 = - z_4, \qquad\beta_1= \beta_2 = 0\,,
\end{align}
and keeps 4 of the 10 scalar fields: $\phi_1= \phi_2=\phi_3$ , $\alpha_1=\alpha_2=\alpha_3$, $\phi_4$ and $\varphi$.
We are interested in solving the BPS equation given in \eqref{eq:10scalarGradientFlowEqst} and we will continue to impose
$\alpha_i=f=0$.

In addition to the $AdS_5$ vacuum solution with vanishing scalar fields and radius $L$,
this model also contains the 
$\mathcal{N}=1$ $SU(3)$ invariant $AdS_4\times\mathbb{R}$ S-fold given in \eqref{n1sfoldsol} (with $\phi_1=\alpha_i=0$), 
and the $\mathcal{N}=4$ $AdS_4\times\mathbb{R}$ S-fold given in \eqref{nfoursfoldsol} (with $\phi_4=-\phi_1$ and $\alpha_i=0$).
In addition, this model contains the GPPZ RG flow of \cite{Girardello:1999bd} that moves from $AdS_5$ with homogeneous mass deformations in the UV to singular (gapped) behaviour in the IR, and this will also appear in our constructions below.

We focus on a one parameter family of RG flow solutions that flow from the vacuum $AdS_5$ in the UV to the 
$\mathcal{N}=4$ $AdS_4\times\mathbb{R}$ S-fold solutions in the IR.
With $\alpha_i=f=0$, the first order BPS equations \eqref{eq:10scalarGradientFlowEqst}
involve 4 real functions: $A,V, \phi_3,\phi_4$, along with $\varphi=k x$. As there is no reality constraint on the $z^A$ fields for this truncation we cannot solve for $V$ algebraically. Taking into account that we can shift the radial coordinate and absorb a constant shift of $A$ by rescaling $(t,y_1,y_2)$ in \eqref{iso21metans}, we deduce that a solution will be specified by specifying two integration constants.
The expansion about the $AdS_5$ vacuum in the UV is given by
\eqref{genscexp}-\eqref{eq:ten_scalar_vev} with 
\begin{align}
\phi_{4(s)} = \frac{kL}{2}, \qquad \phi_{1(v)} = -\phi_{1(s)} \Big( 
\frac{5}{6} \phi_{1(s)}^2 - k L \phi_{1(s)} + \frac{3 k^2L^2}{8}
\Big)\,,
\end{align}
Thus, in addition to the $\varphi=k x$ deformation, there is one independent source, $\phi_{1(s)}$, and 
one independent vev, $\phi_{4(v)}$ and these are the two integration constants that specify an RG flow solution.
In fact for a given $\phi_{1(s)}/k$, our numerical investigations indicate that there is just one RG flow solution that approaches the $\mathcal{N}=4$ $AdS_4\times\mathbb{R}$ S-fold solution in the IR. 

To explain this further, we first notice that if we set $\phi_1=-\phi_4$ then we can recover the RG flow
solutions to the $\mathcal{N}=4$ S-fold solutions that lie
within the $SO(3)\times SO(3)$ invariant truncation which we gave, analytically, 
in section \ref{so3so3trunc}. We saw that these solutions were uniquely specified by $k$ and moreover, that
the truncation had a single irrelevant mode in the IR, with $d=3$ scaling dimension $\Delta=5$, which can be used to move
from the IR to the UV. Within the larger $SO(3)$ invariant truncation we need another irrelevant mode in order to shoot out
from the IR and give rise to a one-parameter family of RG flow solutions with $AdS_5$ in the UV. 
Interestingly, as discussed in appendix \ref{specrta}, the linearised analysis about the $\mathcal{N}=4$ S-fold only reveals an additional marginal mode with $\Delta=3$. Importantly, however, going beyond the linearised analysis we don't find 
an exactly marginal mode but a marginally irrelevant mode instead. This mode can therefore be used to shoot out from the IR
along with the $\Delta=5$ mode. However, the IR expansion is associated with a very slow logarithmic approach to the 
 $\mathcal{N}=4$ S-fold, making a numerical construction of these solutions rather delicate.
 
\begin{figure}[h!]
\centering
{\includegraphics[scale=0.3]{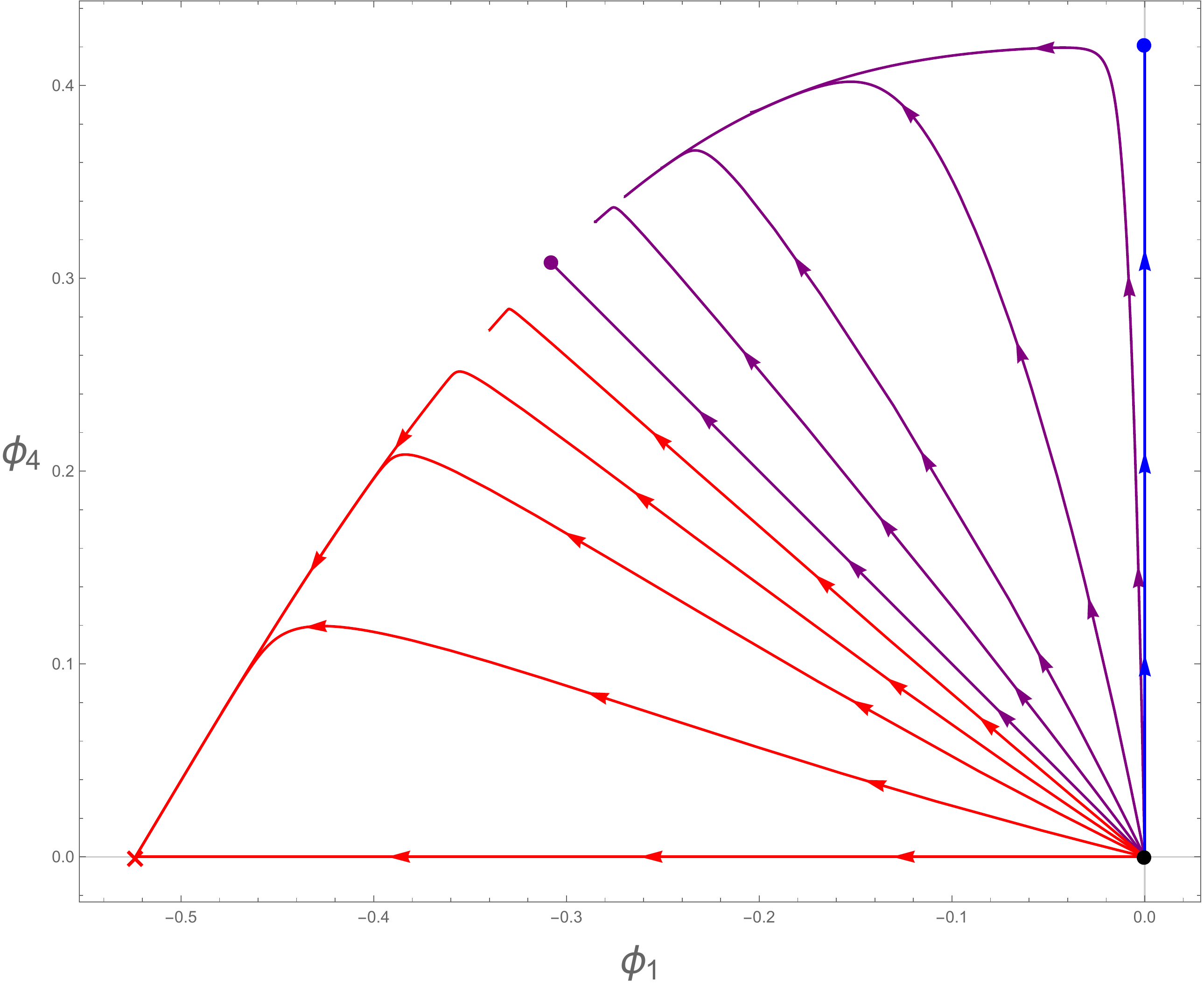}}
\caption{Parametric plots for the RG flows in the $SO(3)$ invariant 4-scalar truncation. The black dot is the $AdS_5$ vacuum, the purple dot 
is the $\mathcal{N}=4$, $SO(3)\times SO(3)$ invariant $AdS_4\times \mathbb{R}$ S-fold solution, while the blue dot is the $\mathcal{N}=1$, $SU(3)$ invariant $AdS_4\times S^1$ S-fold solution. 
The purple line with $\phi_1=-\phi_4$ is the 
$\mathcal{N}=4$ RG flow considered in section \ref{so3so3trunc} and the blue line with $\phi_1=0$
is the $\mathcal{N}=1$ RG flow considered in section \ref{su3}. The horizontal red line is the
homogeneous GPPZ RG flow, which terminates at  a singularity denoted by the red cross.}\label{SO3flows}
\end{figure}

 In practise, in contrast to previous cases we have considered, for this model we numerically integrated the BPS equations
starting from the UV, varying the integration constant $\phi_{4(v)}$ for a given value of $\phi_{1(s)}$. 
While there is some delicate fine tuning involved, we have found strong evidence for a one parameter family of RG flows moving from 
$AdS_5$ in the UV to the $\mathcal{N}=4$ S-fold in the IR, as summarised by the purple lines in figure \ref{SO3flows}. These RG flows
are associated with $\phi_{1(s)}/k\in(0,1/2)$.
Our results also provide strong evidence for the existence of an RG flow that starts at the $\cN=1$ S-fold in the UV and flows to the
$\cN=4$ S-fold in the IR.

Figure \ref{SO3flows} also shows that there is a phase transition at $\phi_{1(s)}/k=1/2$. Indeed, when $\phi_{1(s)}/k>1/2$
we find a different family of RG flows, labelled by the red lines, that start out from $AdS_5$ in the UV and hit a singularity at $\phi_4=0$ and $\phi_1=-\pi/6$. Furthermore, in the limit that $\phi_{1(s)}/k\to\infty$ we approach the horizontal red line, which is precisely the GPPZ RG flow \cite{Girardello:1999bd} associated with homogeneous mass deformations. Finally, this family of solutions
also provide strong evidence for the existence of an RG flow that starts at the $\cN=4$ S-fold in the UV and flows to the
GPPZ singularity, suitably compactified, in the IR.

\section{Discussion}\label{sec:disc}

We have discussed various aspects of S-fold solutions of type IIB string theory of the form
$AdS_4\times{S^1}\times S^5$ which have non-vanishing $SL(2,\mathbb{Z})$ monodromy along the $S^1$ direction
 and are dual to $d=3$ SCFTs with either $\mathcal{N}=4,2$ or 1 supersymmetry. All of our analysis has been within the framework
 of various consistent truncations of maximal $D=5$ gauged supergravity that contain the $D=5$ metric plus some additional scalar fields, including
 the $D=5$ dilaton, $\varphi$. The models have a global shift symmetry $\varphi\to \varphi+ constant$ which plays a key role.
From the $D=5$ point of view the S-fold solutions are all of the form $AdS_4\times\mathbb{R}$ with the dilaton depending linearly on a coordinate parametrising the $\mathbb{R}$ direction. After uplifting to type IIB and acting with $SL(2,\mathbb{R})$ transformations we obtain a larger family of solutions for which the S-folding procedure is carried out.

We presented a new one parameter family of S-fold solutions that are dual to $d=3$ SCFTs preserving $\mathcal{N}=2$ supersymmetry.
The family contains the $\mathcal{N}=4$ S-fold supergravity solution of \cite{Inverso:2016eet,Assel:2018vtq}, as well as the previously constructed 
$\mathcal{N}=2$ $SU(2)\times U(1)$ invariant S-fold solution of \cite{Guarino:2020gfe,Bobev:2020fon}. The new solutions thus correspond to a component of the conformal
manifold containing the $d=3$ SCFT dual to the $\mathcal{N}=4$ S-fold. It would be interesting to elucidate the full conformal manifold; the component preserving $\mathcal{N}=2$ must be K\"ahler with respect to the Zamalodchikov metric, constructed from two point functions of exactly marginal operators,
 and hence there is at least
one more exactly marginal direction to be found.  It should be possible to explicitly identify it 
by enlarging the truncations of maximal $D=5$ gauged supergravity to have additional scalars and/or including $D=5$ vector fields which could have a non-vanishing holonomy along the $\mathbb{R}$ direction.
It would be interesting to know whether the K\"ahler manifold is compact (see \cite{Perlmutter:2020buo} for a recent discussion); 
it seems plausible that it is, with the missing marginal deformation a compact direction which degenerates
at the boundary of the open interval that parametrises the deformations we have found, possibly with a cusp singularity.

We noted that the 10-scalar model has a $\mathbb{Z}_2\times S_4$ symmetry (see (2.9),(2.10) of \cite{Arav:2021tpk}). For all of the solutions we have constructed in this paper one can trivially obtain additional, physically equivalent solutions by acting with these discrete symmetries.
For example, the one parameter family of $\mathcal{N}=2$ S-fold solutions we found are contained within the $U(1)\times U(1)$ invariant sub-truncation of the 10-scalar model, as given in figure \ref{truncdiag}. After acting with elements of the $\mathbb{Z}_2\times S_4$ symmetry of the 10-scalar model (see (2.9),(2.10) of \cite{Arav:2021tpk}), we immediately obtain two other families of  solutions. For
example, instead of $\phi_1=\phi_2$, $\phi_3=-\phi_4$ the other two families have $\phi_2=\phi_3$, $\phi_1=-\phi_4$ and $\phi_1=\phi_3$, $\phi_2=-\phi_4$.

A different one parameter family of 
$\mathcal{N}=2$ S-fold solutions that contains the $SU(2)\times U(1)$ invariant S-fold was found in
\cite{Guarino:2020gfe} and it has recently been shown that this is a compact modulus \cite{Giambrone:2021zvp}. 
This family, which should also be supplemented by another exactly marginal deformation in order to be K\"ahler, intersects
the family we have found just at the $\mathcal{N}=2$ $SU(2)\times U(1)$ invariant S-fold. It will
be very interesting to elucidate the global structure of these different components of the conformal manifold
preserving $\mathcal{N}=2$ supersymmetry as well as any additional components\footnote{Note that the symmetry generating techniques
used in e.g. \cite{Lunin:2005jy,Gauntlett:2005jb,Bobev:2021gza} cannot be used in the present setting because the
solutions do not have a $U(1)\times U(1)$ symmetry that commutes with $R$-symmetry.}. 
One can anticipate that the tools of exceptional generalised geometry, along the lines of 
\cite{Ashmore:2016oug,Ashmore:2018npi,Giambrone:2021zvp}, will be helpful.
Furthermore, it is possible that there are additional components of the conformal manifold
that just preserve $\mathcal{N}=1$ supersymmetry.

We have also constructed type IIB solutions that describe RG flows starting from $AdS_5\times S^5$ in the UV, and then
flow to various
$AdS_4\times{S^1}\times S^5$ S-fold solutions in the IR. In $D=5$ the supergravity solutions flow from $AdS_5$ in the UV, deformed with a linear dilaton plus additional spatially homogeneous fermion and boson mass deformations, down to the $AdS_4\times\mathbb{R}$ S-fold solutions with linear dilatons in the IR. Our construction utilised, in a crucial way, the shift symmetry of the dilaton in the $D=5$ supergravity theory. Indeed
this allows for a Q-lattice type construction \cite{Donos:2013eha} whereby one exactly solves for all of the spatial dependence of the boundary deformations (in this case the linear dilaton), leaving one to solve a set of ODEs that just depend on the holographic radial direction.

In type IIB, after S-folding, these RG flow solutions describe $\mathcal{N}=4$ SYM with a spatially dependent coupling constant $\tau$ that 
traces out a segment of a semi-circular (geodesic) arc in the upper half plane and with fermion mass deformations with spatially dependent phases and constant masses for the bosons.  
The parametrisation of this curve as a function of the $y_3$
coordinate in the boundary theory arises from an $SL(2,\mathbb{R})$ transformation of the uplifted $D=5$ linear dilaton. It is an interesting open question to know whether or not there are similar RG flows if the spatially dependent coupling constant $\tau$ traces 
out the same arc but is parametrised as a function of $y_3$ differently, or more generally, traces out another curve in the
upper half plane\footnote{In the case of $\cN=4,2$ supersymmetry it should be a semi-circular arc\cite{Gaiotto:2008sd,Maxfield:2016lok}.}. In order to preserve supersymmetry we need to have spatially dependent boson and fermion masses as determined from our general analysis in section \ref{sec:2}.
In the case of $\mathcal{N}=4$ supersymmetry the identification of the $d=3$ SCFTs dual to the S-folds made in \cite{Assel:2018vtq} implicitly assumed that the details of the parametrisation $\tau(y_3)$ is not important. To address this 
we could try to generalise our solutions by constructing $D=5$ solutions with $\varphi$ an arbitrary function of $y_3$ and see if we still flow to the same S-fold in the IR. More generally, one will also need to enlarge the supergravity model to include additional fields.
However, such investigations will necessarily involve solving PDEs, since the ansatz for such solutions will lie outside of
the Q-lattice ansatz. Further analysis along these lines could also help clarify the peculiar fact
that $\mathcal{N}=4$ S-folds with a linear dilaton only exist for 
signs of the masses satisfying \eqref{n4cond}, even though this is not required by the supersymmetry analysis.
Note that the absence of such S-fold solutions seems to be related to the fact that there are no associated
$\mathcal{N}=4$ Janus solutions, dual to superconformal interfaces with step function for the coupling constant.

In the case of $\mathcal{N}=4$ deformations, up to signs, the masses are determined by $\tau(y_3)$
- see \eqref{emmsn4}-\eqref{deesn4}. However, for $\mathcal{N}=2,1$ deformations there is more freedom in turning
on independent spatially dependent masses and it will be very interesting to determine when one still flows to the same 
S-fold in the IR. Naively one might expect that spatially dependent masses will die out along the RG flow and become irrelevant in the IR.
However, in the case of $\mathcal{N}=1$ deformations, we expect the full story to be significantly more involved. The reason
for this is that in addition to the $\mathcal{N}=1$ S-fold solutions with a linear $D=5$ dilaton that we have been discussing, 
there are also infinite examples of S-fold solutions with the $D=5$ dilaton a linear plus periodic (LPP) function of the spatial direction \cite{Arav:2021tpk}. In these LPP solutions the remaining scalar fields are all periodic functions of the spatial direction and the
S-folding procedure is carried out over one or more periods that are set by the solution.
It is a fascinating open issue to determine which UV deformations of $\mathcal{N}=4$ SYM that we studied in section \ref{sec:2}
can flow to this array of
$d=3$ SCFTs. Another closely related topic is to determine which UV deformations of $\mathcal{N}=4$ SYM can flow to
the purely periodic $AdS_4\times S^1\times S^5$ solution (not S-folded) that was constructed in \cite{Arav:2020obl}.

\vskip 1 cm
{\bf Note Added in Print:}
Several days after posting our paper to the arXive, \cite{Bobev:2021yya} appeared
which has overlap with some of the results of this paper.

\subsection*{Acknowledgments}
We thank Antoine Bourget, Matthew Cheung and Amihay Hanany for helpful discussions.
This work was supported by STFC grant ST/T000791/1. JPG is supported as a Korea Institute for Advanced Studies (KIAS) Scholar and as a Visiting Fellow at the Perimeter Institute for Theoretical Physics. 
The work of CR is funded by a Beatriu de Pin\'os Fellowship.

\appendix

\section{Rewriting the 10-scalar BPS equations as a gradient flow}\label{derivbps}
We start by recalling the BPS equations derived in \cite{Arav:2021tpk} for $D=5$ configurations that preserve 
$ISO(2,1)$ symmetry. The metric is assumed to be of the form
\begin{align}\label{iso21metansap}
ds^2 = e^{2A}(dt^2 - dy_1^2 - dy_2^2) - e^{2V} (dx^2 + dr^2)\,,
\end{align}
with $A,V$ and all scalars functions of $(r,x)$ only. For simplicity we will 
write down the BPS equations with, in the notation of  \cite{Arav:2020obl}, $\kappa=+1$
which corresponds to Killing spinors satisfying a projection condition with a specific sign.
We define the complex coordinate
$w=r-ix$ and also a one-form $B$ given by
\begin{align}
B\equiv bdw\,,\qquad b\equiv \frac{1}{6}e^{i\xi+V+\mathcal{K}/2}{\mathcal{W}}\,.
\end{align}

The BPS equations from section 4 of \cite{Arav:2020obl} are given by
\begin{align}\label{parelbeetext}
\partial A&=B\,,\nn
\bar \partial B&= - \mathcal{F}B\wedge \bar B\,,\nn
\bar\partial z^A&=-\frac{3}{2}(\bar{\mathcal{W}})^{-1}\mathcal{K}^{\bar B A}\nabla_{\bar B}\bar{\mathcal{W}}\bar B\,,\nn
\bar\partial \beta_1&=-\frac{1}{4}(\bar{\mathcal{W}})^{-1}\partial_{\beta_1}\bar{\mathcal{W}}\bar B\,,\nn
\bar\partial \beta_2&=-\frac{3}{4}(\bar{\mathcal{W}})^{-1}\partial_{\beta_2}\bar{\mathcal{W}}\bar B\,.
\end{align}
where $\mathcal{F}$ is a real quantity just depending on $\mathcal{W}$, $\mathcal{K}$ given by
\begin{equation}\label{ceffdef}
\mathcal{F} \equiv 1-\frac{3}{2}\frac{1}{|\mathcal{W}|^2}\nabla_A\mathcal{W}\mathcal{K}^{A\bar B}\nabla_{\bar B}\bar{\mathcal{W}}
-\frac{1}{4}|\partial_{\beta_1}\log \mathcal{W}|^2
-\frac{3}{4}|\partial_{\beta_2}\log \mathcal{W}|^2\,,
\end{equation}
and $\partial$, $\bar \partial$ are the holomorphic and anti-holomorphic exterior derivatives.
In these equations $\partial_{\beta_i}\bar{\mathcal{W}}$ is an ordinary real derivative, in contrast to below.

We now make the ansatz that the only dependence of the scalar fields on the $x$ direction is via the $D=5$ dilaton:
$\varphi= kx+f(r)$ with all other scalars just depending on $r$. Thus, we assume
\begin{equation}
\partial_x z^A = k l^A,\qquad
\partial_x A = \partial_x V = \partial_x \beta_i = 0 \,,
\end{equation}
where $l^A$ is the holomorphic Killing vector on the scalar manifold corresponding to the dilaton shift symmetry \eqref{dilshift}.
The flow equations then take the following form:
\begin{align}\label{radbpseqs}
& \partial_r A = 2 b\,,\nn
& \partial_r z^A = -3 K^{\bar{B}A} \partial_{\bar{B}}\widetilde{\mathcal{K}}b + ik l^A = -K^{\bar{B}A} \left( 3\partial_{\bar{B}} \widetilde{\mathcal{K}} b - k \partial_{\bar{B}}\mu \right) \,,\nn
& \partial_r \beta_1 = - \frac{1}{2} \partial_{\bar{\beta_1}} \widetilde{\mathcal{K}} b \,,\nn
& \partial_r \beta_2 = - \frac{3}{2} \partial_{\bar{\beta_2}} \widetilde{\mathcal{K}} b \,,\nn
&b^{-2}\partial_r b = 2 \mathcal{F}\,,
\end{align}
where $\mu$ is the moment map corresponding to the Killing vector $l^A$ given in \eqref{mommapcon} and
\begin{align}
& \mathcal{F} \equiv 1 - \frac{3}{2} K^{\bar{B} A} \partial_A \widetilde{\mathcal{K}} \partial_{\bar{B}} \widetilde{\mathcal{K}} - \frac{1}{4} |\partial_{\beta_1} \mathcal{\widetilde{K}}|^2 - \frac{3}{4} |\partial_{\beta_2} \mathcal{\widetilde{K}}|^2 \,.
\end{align}
Note that here we have switched notation: $\partial_{\bar{\beta_i}}$ is now shorthand for taking $\beta_i$ to be a complex field in the superpotential such that $\mathcal{W}$ is a function of $\beta_i$ and $\overline{\mathcal{W}}$ of $\bar{\beta_i}$, then taking the appropriate derivative and finally setting $\beta_i$ to be real.
Thus, $ \partial_{\bar{\beta_1}} \mathcal{W}\equiv0$ and $\partial_{\bar{\beta_1}} \overline{\mathcal{W}}$ is given by the ordinary real derivative $\partial_{{\beta_1}} \overline{\mathcal{W}}$.
We also observe from the first equation that $b$ is real and hence we can write
\begin{align}
b=\frac{1}{6}e^{V+\widetilde{\mathcal{K}}/2}\,,
\end{align}
where we fixed a sign without loss of generality.

Using these equations, one can easily derive the following equations:
\begin{align}\label{beeeq}
& \partial_r \widetilde{\mathcal{K}} = 4 (\mathcal{F}-1) b+ k \left( i l^A \partial_A \widetilde{\mathcal{K}} - i l^{\bar{A}} \partial_{\bar{A}} \widetilde{\mathcal{K}} \right) =  4  (\mathcal{F}-1) b + 2k \mu \,, \nn
&\partial_r V = 2 b- k \mu\,.
\end{align}
It is now straightforward to obtain the RG flow equations as written in 
\eqref{eq:10scalarGradientFlowEqst}.

\section{BPS spectrum for S-folds}\label{specrta}

The S-fold solutions with linear dilatons solve the BPS equations \eqref{eq:10scalarGradientFlowEqst} with $A= +r/\ell$ and $V$ constant.
The $D=5$ metric is given by \eqref{fixedptsol} and the canonically normalised $D=4$ Einstein metric is given by \eqref{4dads4}.
Here we calculate the spectrum of BPS scalar fluctuations about these S-folds by analysing the linearised fluctuations acting
on the set $\Phi=\{z^A,\beta_i,V\}$. These naturally fall into two categories: BPS non-normalisable modes (``sources"), corresponding to a fluctuation of the form $\delta \Phi \sim e^{-(3-\Delta) r/\ell}$, and BPS normalisable modes (``vevs") corresponding to fluctuations of the form $\delta \Phi \sim e^{-\Delta r/\ell}$, where the conformal boundary for the S-folds is located at $r\to+\infty$. In these expressions, generically
$\Delta$ is the scaling dimension of the operator in the dual $d=3$ SCFT that is associated with a source or vev, but in the window 
$1\le \Delta\le 2$ there is the usual ambiguity of which quantisation conditions are imposed (restricted by supersymmetry in the present case).

Since the S-folds solutions with a linear dilaton are invariant under the $\mathbb{Z}_2$ symmetry $z^A \leftrightarrow -\zb^A$,
we can label operators by their charge under the action of $z^A \rightarrow - \bar{z}^A$ and this is denoted with a 
$\pm$ subscript on the $\Delta$'s in the tables below. A more careful study determining other quantum numbers for these modes and how they arrange themselves into partial supermultiplets is left to future work, but can be partially inferred from the tables below.

Note that there is a universal mode with $\Delta=3_-$ which corresponds to the shift $\varphi\to \varphi+ constant$.
After a Scherk-Schwarz reduction this mode corresponds to gauge transformation in the $D=4$ theory and does
not correspond to a physical marginal mode in the dual $d=3$ SCFT. We have not included this mode in the tables.

For each of the S-fold solutions we give the BPS spectrum for the 10-scalar truncation as well as for each of the subtruncations
that contains it (see figure \ref{truncdiag}). The latter is helpful for keeping track of which modes are active in the RG flow solutions that we construct in the main text.

\subsection{$\cN=1$ $SU(3)$ invariant S-fold}
The $\cN=1$ $SU(3)$ invariant S-fold solution is given by
\begin{align}
\phi_1&=\phi_2=\phi_3=0\,,\qquad \beta_1=\beta_2=0\,,\nn
\cos \phi_4&=\sqrt{\frac{{5}}{{6}}},\quad k \ell = \frac{\sqrt{5}}{2}\,,\quad e^V = \frac{5L}{6\ell}\,,
\end{align}
and the BPS spectrum is given in table \ref{bpsn1}.
\begin{table}[htp]
\begin{center}
\begin{tabular}{|c|c|c|}
\hline 
$\cN=1$ $SU(3)$& & \\
\hline
Truncation& Source & Vev \\ 
\hline
10-scalar & $(\frac{5}{3})_+\times 3,~(2+\sqrt{6})_+$
& $2_+\times 2,~ (1+\sqrt{6})_+,~(\frac{5}{3})_-\times 3 $
 \\ \hline
 $SU(2)$ \text{5-scalar}&
 $ (\frac{5}{3})_+,~(2+\sqrt{6})_+$  
 &
 $(\frac{5}{3})_-,~2_+,~(1+\sqrt{6})_+$
 \\ \hline
 $SO(3)$ 4-scalar& 
 $(\frac{5}{3})_+,~(2+\sqrt{6})_+$
 &
 $(\frac{5}{3})_-,~(1+\sqrt{6})_+$ \\ 
\hline
 $SU(3)$ 2-scalar & 
 $(2+\sqrt{6})_+$
 &
 $(1+\sqrt{6})_+$ \\ 
\hline
\end{tabular}
\end{center}
\caption{BPS modes and multiplicities for the $\cN=1$ $SU(3)$ invariant S-fold.\label{bpsn1}}
\end{table}%

\subsection{$\cN=2$ $SU(2)\times U(1)$ invariant S-fold}
The $\cN=2$ $SU(2)\times U(1)$ invariant S-fold solution is given by
\begin{align}
\phi_1&=\phi_2=0,\qquad  \beta_1=0,\nn
\phi_3&=-\phi_4 = -\frac{\pi}{8}, \quad  \beta_2 = -\frac{\log 2}{12} ,\quad k \ell = 1, \quad  e^V = \frac{L}{2^{1/3}\ell}\,,
\end{align}
and the BPS spectrum is given in table \ref{bpsn2app}. The marginal mode with $\Delta=3_+$ in the 
$U(1)\times U(1)$ 6-scalar and the 10-scalar truncation is the marginal mode that takes one along
the $\mathcal{N}=2$ conformal manifold discussed in section \ref{appbfam}.
\begin{table}[htp]
\begin{center}
\begin{tabular}{|c|c|c|}
\hline 
$\cN=2$ $SU(2)\times U(1)$ & & \\
\hline
Truncation& Source & Vev \\ 
\hline
10-scalar &
$2_+,~2_-,~3_+,~(\frac{3+\sqrt{17}}{2})_+,~( \frac{5+\sqrt{17}}{2})_+$ & 
$2_+,~2_-,~( \frac{1+\sqrt{17}}{2})_+,~( \frac{3+\sqrt{17}}{2})_+$ \\
\hline
$U(1)\times U(1)$ 6-scalar & 
$2_-,~3_+,~( \frac{5+\sqrt{17}}{2})_+$ &
$( \frac{1+\sqrt{17}}{2})_+$ \\ \hline
$SU(2)$ 5-scalar&
$2_-,~( \frac{3+\sqrt{17}}{2})_+,~( \frac{5+\sqrt{17}}{2})_+$ &
$( \frac{1+\sqrt{17}}{2})_+,~( \frac{3+\sqrt{17}}{2})_+$ 
\\ \hline
$SU(2)\times U(1)$ 4-scalar&
$( \frac{5+\sqrt{17}}{2})_+$ &
$( \frac{1+\sqrt{17}}{2})_+$ 
\\ \hline
\end{tabular}
\end{center}
\caption{BPS modes and multiplicities for the $\cN=2$ $SU(2)\times U(1)$ invariant S-fold.\label{bpsn2app}}
\end{table}%

\subsection{$\cN=4$ $SO(3)\times SO(3)$ invariant S-fold}
The $\cN=4$ $SO(3)\times SO(3)$ invariant S-fold solution is given by
\begin{align}
\beta_1 &=\beta_2 = 0,\nn
\phi_1 &= \phi_2=\phi_3=-\phi_4 = - \mathrm{arccot}(\sqrt{2})/2,\quad ~ k \ell = \frac{1}{\sqrt{2}},\quad e^V = \frac{L}{\sqrt{2} \ell}\,,
\end{align}
and the BPS spectrum is given in table \ref{bpsn4}.
In the $U(1)\times U(1)$ invariant 6-scalar truncation the marginal mode with $\Delta=3_+$ is the mode
that takes one along the $\mathcal{N}=2$ conformal manifold discussed in section \ref{appbfam}.
In the 10-scalar truncation there are two further marginal modes with $\Delta=3_+$ that correspond to 
two more $\mathcal{N}=2$ conformal manifolds emanating from the $\mathcal{N}=4$ S-fold that arise from the
action of the $\mathbb{Z}_2\times S_4$ symmetry of the 10-scalar model.
In the $SO(3)$ invariant 4-scalar truncation we 
appear to have a $\Delta=3_+$ exactly marginal operator. However, this flat direction does not survive to nonlinear order, as it corresponds to turning on a linear combination of two of the three exactly marginal modes of the 10-scalar model. Since this mode is not exactly marginal, we have
labelled this mode with an asterisk.
\begin{table}[h!tp]
\begin{center}
\begin{tabular}{|c|c|c|}
\hline
$\cN=4$ $SO(3)\times SO(3)$ &  & \\
\hline
Truncation& Source & Vev \\ 
\hline
10 scalar & 
$2_-,~3_+\times 3,~5_+$ & 
$3_+\times 3$ \\ 
\hline
$U(1)\times U(1)$ 6-scalar  &
$2_-,~3_+,~5_+$   &
$3_+$ 
\\ \hline
$SO(3)$ 4-scalar &
$2_-,~3_+^*,~5_+$   &
$3_+$ 
\\ \hline
$SO(3)\times SO(3)$ 3-scalar&
$2_-,~5_+$   & $-$
\\ \hline
\end{tabular}
\end{center}
\caption{BPS modes and multiplicities for the $\cN=4$ $SO(3)\times SO(3)$ invariant S-fold. \label{bpsn4}}
\end{table}%

 \subsection{Family of $\mathcal{N}=2$ S-folds}\label{appbfam}
 
\begin{figure}[htbp]
\begin{center}
\includegraphics[scale=.7]{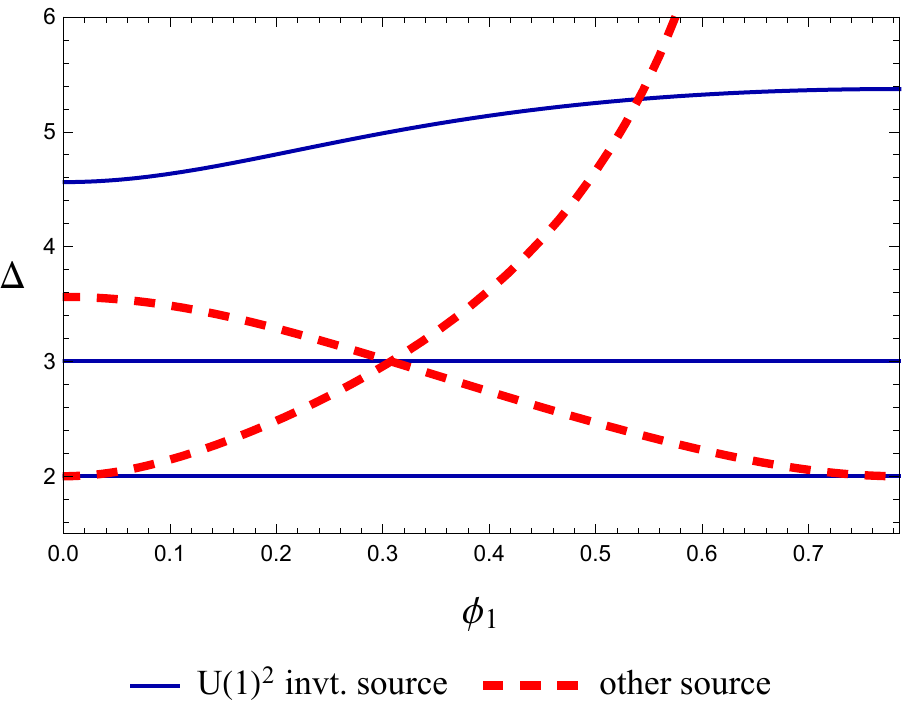} \qquad
\includegraphics[scale=.7]{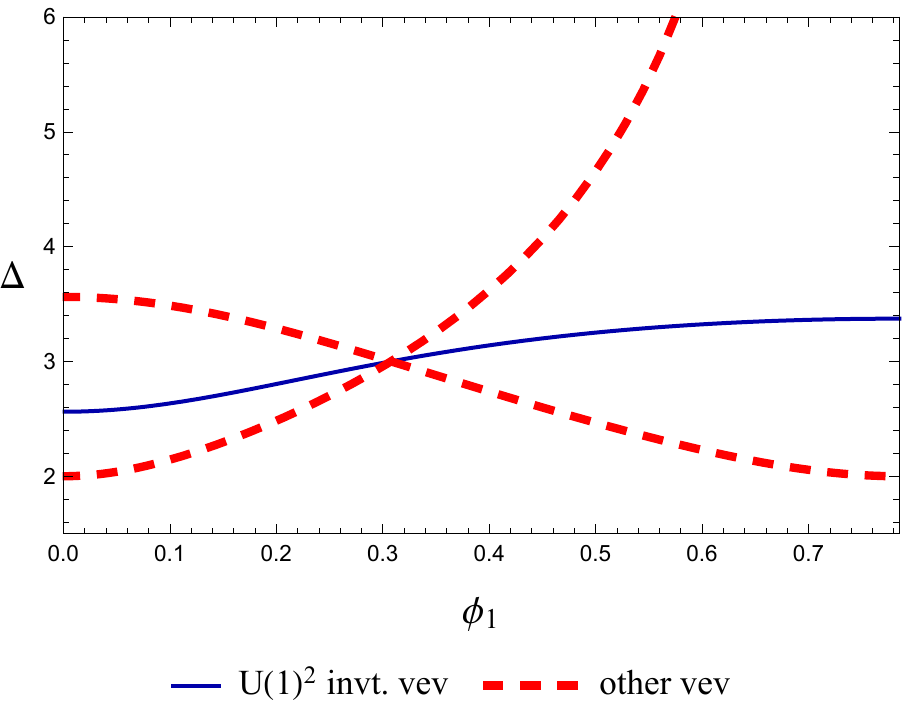} 
\caption{Dimensions of BPS modes as we move on the conformal manifold \eqref{confman}, labelled by the value of $\phi_1$, and we have only shown $\phi_1\ge 0$. The $\cN=2$ $SU(2)\times U(1)$ invariant S-fold is located at $\phi_1=0$ and 
the $\cN=4$ S-fold is located at $\phi_1= \text{arccos}(1/3)/4\sim0.31$. 
The left plot gives the dimensions of operators which admit BPS sources, and the right those which admit BPS vevs. The blue curves are the BPS modes in the $U(1)\times U(1)$ invariant truncation and the red dashed curves
are the remaining modes in the 10-scalar truncation.}\label{bpsfamfig}
\label{fig:dims}
\end{center}
\end{figure}

The one-parameter family of $\mathcal{N}=2$ S-folds that can be constructed in the $U(1)\times U(1)$ invariant 6-scalar truncation
is given in \eqref{confman}. The solutions can be labelled by the value of $\phi_1\in (-\pi/4,\pi/4)$ and are symmetric under $\phi_1\to-\phi_1$.
The BPS spectrum is given in table \ref{bpsfam} and we have displayed them graphically in figure \ref{bpsfamfig}.
As we move on the conformal manifold, the scaling dimensions change. In particular, as we cross the $\cN=4$ point at $\phi_1= \text{arccos}(1/3)/4$ we see operators changing from relevant to irrelevant and vice-versa, as well as the appearing of the three marginal modes at
the $\cN=4$ point itself. Note also that one of the scaling dimensions of the BPS sources diverges at the boundary of the conformal manifold at $\phi_1=\pm \pi/4$.
\begin{table}[htp]
\begin{center}
\begin{tabular}{|c|c|c|}
\hline
$\cN=2$ $U(1)\times U(1)$ & & \\
\hline
Truncation& Source & Vev \\ 
\hline
10-scalar &
$2_-,~
(\frac{3}{2}+\sqrt{\frac{16-15c^2}{4c^2}})_+,
$ & 
$
(\frac{3}{2}+\sqrt{\frac{16-15c^2}{4c^2}})_+,
$ \\
&
$
(\frac{3}{2}+\sqrt{\frac{2+15 c^2}{4(2-c^2)}})_+,$ & 
$ 
(\frac{1}{2} + \sqrt{\frac{66-49c^2}{4(2-c^2)}})_+ $
\\
& $
3_+,~
(\frac{5}{2} + \sqrt{\frac{66-49c^2}{4(2-c^2)}})_+,$ & $
(\frac{3}{2}+\sqrt{\frac{2+15 c^2}{4(2-c^2)}})_+$ 
\\ \hline
$U(1)\times U(1)$ 6-scalar& 
$2_-,~3_+,~(\frac{5}{2} + \sqrt{\frac{66-49c^2}{4(2-c^2)}})_+$ & 
$(\frac{1}{2} + \sqrt{\frac{66-49c^2}{4(2-c^2)}})_+$
 \\ \hline
\end{tabular}
\end{center}
\caption{BPS modes for the one-parameter family of $\cN=2$ $U(1)\times U(1)$ invariant S-folds, where $c = \cos(2\phi_1)$.\label{bpsfam}}
\end{table}%

\section{Enhancement of supersymmetry}\label{appc}

The BPS equations \eqref{radbpseqst} that we have been studying for the 10-scalar model arise first from a projection on to a particular $\mathcal{N}=2$ subsector of the full $\mathcal{N}=8$ gauged supergravity theory, as explained in \cite{Bobev:2016nua},
parametrised by two symplectic pairs of Majorana spinors. We then further restrict to 
$D=5$ configurations that preserve $ISO(2,1)$ symmetry, as discussed in \cite{Arav:2020obl}, which generically imposes two projection
conditions leading to solutions preserving $\mathcal{N}=1$ Poincar\'e supersymmetry in $d=3$. For the $AdS_4\times\mathbb{R}$ S-fold solutions there is, of course,
the additional superconformal supersymmetries.

For certain subtruncations of the 10-scalar model, there is an enhancement of supersymmetry which arises from additional supersymmetries of the 
$\mathcal{N}=8$ gauged supergravity theory. To see this we can consider the tensor $W_{ab}$ which appears in the fermion variations of the 
$\mathcal{N}$=8 theory. Using the notation of \cite{Gunaydin:1985cu}, in the Einstein scalar sector these variations are given by
\begin{align}
\delta\psi_{\mu\,a}  & =  D_\mu \epsilon_a-\frac{g}{6} W_{ab}\gamma_\mu \epsilon^b\,,\nn
\delta\chi_{abc} & = \sqrt{2}\gamma^\mu P_{\mu\,abcd}\epsilon^d-\frac{g}{\sqrt{2}}A_{dabc}\epsilon^d\,,
\end{align}
where $a=1,\dots,8$ is a $USp(8)$ index. 
As discussed in \cite{Bobev:2016nua}, the 10-scalar model $W_{ab}$ has four sets of complex conjugate eigenvalues 
$\{\lambda_{1,2,3,4},\bar \lambda_{1,2,3,4}\}$
and generically,
only one of them has a holomorphic structure and can be written as 
\begin{align}
\lambda_4=e^{\mathcal{K}/2}\mathcal{W}\,,
\end{align}
with $\mathcal{K}$ and $\mathcal{W}$ given by \eqref{kpot} and 
\eqref{superpotlike}, respectively. 
By restricting $\epsilon^a$ to lie within 
the eigenspace spanned by the eigenvectors with eigenvalues $\lambda_4$ and $\bar\lambda_4$
we obtain the supersymmetry variations for the $\mathcal{N}=2$ subsector mentioned above.

We will now use the fact that the equations of motion for the 10-scalar model are invariant under an additional set of discrete $S_4$ symmetries
given\footnote{Note that the $S_4$ discussed in appendix B of \cite{Arav:2021tpk} is related to this one after combining
with the $S_4$ symmetries mentioned in section \ref{sixscalar} (see 
(2.8)-(2.10) of \cite{Arav:2021tpk})} by \cite{Arav:2021tpk}
\begin{align}\label{secretS4}
\{z^2 &\leftrightarrow + \zb^3
\quad\Leftrightarrow\quad \phi_1 \leftrightarrow -\phi_4\,,\alpha_2 \leftrightarrow \alpha_3\}\,,\quad
\beta_1 \rightarrow +\frac{1}{2}(\beta_2 - \beta_1)\,,\quad \beta_2 \rightarrow \frac{1}{2}(3 \beta_1+\beta_2)\,,\nn
\{z^3 &\leftrightarrow - \zb^4
\quad\Leftrightarrow\quad \phi_2 \leftrightarrow -\phi_4\,,\alpha_1 \leftrightarrow \alpha_3\}\,,\quad
\beta_1 \rightarrow -\frac{1}{2}(\beta_1 + \beta_2)\,,\quad \beta_2 \rightarrow \frac{1}{2}(\beta_2-3\beta_1)\,,\nn
\{z^2 &\leftrightarrow - \zb^4
\quad\Leftrightarrow\quad \phi_3 \leftrightarrow -\phi_4\,,\alpha_1 \leftrightarrow \alpha_2\}\,,\quad
 \beta_2 \rightarrow - \beta_2
\,.
\end{align}
In contrast to the $\mathbb{Z}_2\times S_4$ symmetries mentioned in section \ref{sixscalar} (see 
(2.8)-(2.10) of \cite{Arav:2021tpk}) 
these symmetries do not preserve the BPS equations but instead transform the supercharges into each other; they are discrete $R$-symmetry transformations that generalise (3.3) of \cite{Khavaev:2000gb} to include the $\alpha_i$ scalars. 
Acting on a solution to the BPS equations with these $S_4$ transformations, one obtains another solution to the equations of motion  of the 10-scalar model. This will then be a supersymmetric solution of $D=5$ maximal gauged supergravity, but preserving a different set of Killing spinors. This also implies that if one has a supersymmetric solution in a consistent truncation that is invariant under
one of the generators \eqref{secretS4}, then it will automatically preserve twice as much supersymmetry.

For example, the $U(1)\times U(1)$ invariant 6-scalar truncation is invariant under the last transformation in \eqref{secretS4} and hence solutions to the BPS equations that we have been considering in this paper generically preserve $\cN=2$ supersymmetry. Similarly, the $SO(3)\times SO(3)$ invariant truncation is invariant under the full $S_4$ and is associated with the preservation of $\cN=4$ supersymmetry.

The above comments imply that this $S_4$ transforms the eigenvalues of the $W$-tensor into each other and furthermore, in a truncation of the 10-scalar model which is invariant under one of the generators of the discrete $S_4$ transformations \eqref{secretS4} we must find that pairs of the $\lambda_i$ become equal. By explicit computation, for the $U(1)\times U(1)$ invariant 6-scalar truncation, using the results of appendix A of \cite{Arav:2021tpk},
we find that the complex eigenvalues take the form
\begin{align}\label{equall3l4}
\lambda_3&=\lambda_4=e^{\mathcal{K}/2}\mathcal{W}\,,
\end{align}
with
\begin{align}
e^{\mathcal{K}/2}\mathcal{W} = \frac{1}{2}e^{-4\beta_1}\frac{1+2z_1z_2+z_2^2-z_3(z_1+2z_2+z_1 z_2^2)-2e^{6\beta_1}(z_2^2-1)(1+z_1 z_3)}{(z_2^2 - 1)\sqrt{1-|z_1|^2}\sqrt{1-|z_3|^2}},
\end{align}
and
\begin{align}\label{equall1l2}
\lambda_1=\lambda_2= \frac{1}{2}e^{-4\beta_1}\frac{1+2z_1z_2+z_2^2-\bar{z}_3(z_1+2z_2+z_1 z_2^2)-2e^{6\beta_1}(z_2^2-1)(1+z_1 \bar{z}_3)}{(z_2^2 - 1)\sqrt{1-|z_1|^2}\sqrt{1-|z_3|^2}}\,.
\end{align}
If we further restrict to the $SO(3)\times SO(3)$ invariant 3-scalar model by setting $z_2=z_3=\bar z_3$ and $\beta_1=0$ we find that
$\lambda_1=\lambda_2=\lambda_3=\lambda_4=e^{\mathcal{K}/2}\mathcal{W}$.


\begin{thebibliography}{10}

\bibitem{Inverso:2016eet}
G.~Inverso, H.~Samtleben, and M.~Trigiante, ``{Type II supergravity origin of
  dyonic gaugings},'' \href{http://dx.doi.org/10.1103/PhysRevD.95.066020}{{\em
  Phys. Rev. D} {\bfseries 95} no.~6, (2017) 066020},
  \href{http://arxiv.org/abs/1612.05123}{{\ttfamily arXiv:1612.05123
  [hep-th]}}.

\bibitem{Assel:2018vtq}
B.~Assel and A.~Tomasiello, ``{Holographic duals of 3d S-fold CFTs},''
  \href{http://dx.doi.org/10.1007/JHEP06(2018)019}{{\em JHEP} {\bfseries 06}
  (2018) 019}, \href{http://arxiv.org/abs/1804.06419}{{\ttfamily
  arXiv:1804.06419 [hep-th]}}.

\bibitem{Bobev:2019jbi}
N.~Bobev, F.~F. Gautason, K.~Pilch, M.~Suh, and J.~Van~Muiden, ``{Janus and
  J-fold Solutions from Sasaki-Einstein Manifolds},''
  \href{http://dx.doi.org/10.1103/PhysRevD.100.081901}{{\em Phys. Rev. D}
  {\bfseries 100} no.~8, (2019) 081901},
  \href{http://arxiv.org/abs/1907.11132}{{\ttfamily arXiv:1907.11132
  [hep-th]}}.

\bibitem{Guarino:2019oct}
A.~Guarino and C.~Sterckx, ``{S-folds and (non-)supersymmetric Janus
  solutions},'' \href{http://dx.doi.org/10.1007/JHEP12(2019)113}{{\em JHEP}
  {\bfseries 12} (2019) 113}, \href{http://arxiv.org/abs/1907.04177}{{\ttfamily
  arXiv:1907.04177 [hep-th]}}.

\bibitem{Guarino:2020gfe}
A.~Guarino, C.~Sterckx, and M.~Trigiante, ``{$\mathcal{N}=2$ supersymmetric
  S-folds},'' \href{http://dx.doi.org/10.1007/JHEP04(2020)050}{{\em JHEP}
  {\bfseries 04} (2020) 050}, \href{http://arxiv.org/abs/2002.03692}{{\ttfamily
  arXiv:2002.03692 [hep-th]}}.

\bibitem{Bobev:2020fon}
N.~Bobev, F.~F. Gautason, K.~Pilch, M.~Suh, and J.~van Muiden, ``{Holographic
  interfaces in $ \mathcal{N} $ = 4 SYM: Janus and J-folds},''
  \href{http://dx.doi.org/10.1007/JHEP05(2020)134}{{\em JHEP} {\bfseries 05}
  (2020) 134}, \href{http://arxiv.org/abs/2003.09154}{{\ttfamily
  arXiv:2003.09154 [hep-th]}}.

\bibitem{Arav:2021tpk}
I.~Arav, K.~C.~M. Cheung, J.~P. Gauntlett, M.~M. Roberts, and C.~Rosen, ``{A
  new family of $AdS_4$ S-folds in type IIB string theory},''
  \href{http://dx.doi.org/10.1007/JHEP05(2021)222}{{\em JHEP} {\bfseries 05}
  (2021) 222}, \href{http://arxiv.org/abs/2101.07264}{{\ttfamily
  arXiv:2101.07264 [hep-th]}}.

\bibitem{DHoker:2006qeo}
E.~D'Hoker, J.~Estes, and M.~Gutperle, ``{Interface Yang-Mills, supersymmetry,
  and Janus},'' \href{http://dx.doi.org/10.1016/j.nuclphysb.2006.07.001}{{\em
  Nucl. Phys.} {\bfseries B753} (2006) 16--41},
\href{http://arxiv.org/abs/hep-th/0603013}{{\ttfamily arXiv:hep-th/0603013
  [hep-th]}}.

\bibitem{Gaiotto:2008sa}
D.~Gaiotto and E.~Witten, ``{Supersymmetric Boundary Conditions in N=4 Super
  Yang-Mills Theory},'' \href{http://dx.doi.org/10.1007/s10955-009-9687-3}{{\em
  J. Statist. Phys.} {\bfseries 135} (2009) 789--855},
\href{http://arxiv.org/abs/0804.2902}{{\ttfamily arXiv:0804.2902 [hep-th]}}.

\bibitem{Gaiotto:2008sd}
D.~Gaiotto and E.~Witten, ``{Janus Configurations, Chern-Simons Couplings, And
  The theta-Angle in N=4 Super Yang-Mills Theory},''
  \href{http://dx.doi.org/10.1007/JHEP06(2010)097}{{\em JHEP} {\bfseries 06}
  (2010) 097},
\href{http://arxiv.org/abs/0804.2907}{{\ttfamily arXiv:0804.2907 [hep-th]}}.

\bibitem{Terashima:2011qi}
Y.~Terashima and M.~Yamazaki, ``{SL(2,R) Chern-Simons, Liouville, and Gauge
  Theory on Duality Walls},''
  \href{http://dx.doi.org/10.1007/JHEP08(2011)135}{{\em JHEP} {\bfseries 08}
  (2011) 135}, \href{http://arxiv.org/abs/1103.5748}{{\ttfamily arXiv:1103.5748
  [hep-th]}}.

\bibitem{Ganor:2014pha}
O.~J. Ganor, N.~P. Moore, H.-Y. Sun, and N.~R. Torres-Chicon, ``{Janus
  configurations with SL(2,$\mathbb{Z}$)-duality twists, strings on mapping
  tori and a tridiagonal determinant formula},''
  \href{http://dx.doi.org/10.1007/JHEP07(2014)010}{{\em JHEP} {\bfseries 07}
  (2014) 010}, \href{http://arxiv.org/abs/1403.2365}{{\ttfamily arXiv:1403.2365
  [hep-th]}}.

\bibitem{Gang:2015wya}
D.~Gang, N.~Kim, M.~Romo, and M.~Yamazaki, ``{Aspects of Defects in 3d-3d
  Correspondence},'' \href{http://dx.doi.org/10.1007/JHEP10(2016)062}{{\em
  JHEP} {\bfseries 10} (2016) 062},
  \href{http://arxiv.org/abs/1510.05011}{{\ttfamily arXiv:1510.05011
  [hep-th]}}.

\bibitem{Garozzo:2018kra}
I.~Garozzo, G.~Lo~Monaco, and N.~Mekareeya, ``{The moduli spaces of $S$-fold
  CFTs},'' \href{http://dx.doi.org/10.1007/JHEP01(2019)046}{{\em JHEP}
  {\bfseries 01} (2019) 046}, \href{http://arxiv.org/abs/1810.12323}{{\ttfamily
  arXiv:1810.12323 [hep-th]}}.

\bibitem{Garozzo:2019hbf}
I.~Garozzo, G.~Lo~Monaco, and N.~Mekareeya, ``{Variations on $S$-fold CFTs},''
  \href{http://dx.doi.org/10.1007/JHEP03(2019)171}{{\em JHEP} {\bfseries 03}
  (2019) 171}, \href{http://arxiv.org/abs/1901.10493}{{\ttfamily
  arXiv:1901.10493 [hep-th]}}.

\bibitem{Garozzo:2019ejm}
I.~Garozzo, G.~Lo~Monaco, N.~Mekareeya, and M.~Sacchi, ``{Supersymmetric
  Indices of 3d $S$-fold SCFTs},''
  \href{http://dx.doi.org/10.1007/JHEP08(2019)008}{{\em JHEP} {\bfseries 08}
  (2019) 008}, \href{http://arxiv.org/abs/1905.07183}{{\ttfamily
  arXiv:1905.07183 [hep-th]}}.

\bibitem{Garozzo:2019xzi}
I.~Garozzo, N.~Mekareeya, and M.~Sacchi, ``{Duality walls in the 4d $
  \mathcal{N} $ = 2 SU(N) gauge theory with $2N$ flavours},''
  \href{http://dx.doi.org/10.1007/JHEP11(2019)053}{{\em JHEP} {\bfseries 11}
  (2019) 053}, \href{http://arxiv.org/abs/1909.02832}{{\ttfamily
  arXiv:1909.02832 [hep-th]}}.

\bibitem{Garozzo:2020pmz}
I.~Garozzo, N.~Mekareeya, M.~Sacchi, and G.~Zafrir, ``{Symmetry enhancement and
  duality walls in 5d gauge theories},''
  \href{http://dx.doi.org/10.1007/JHEP06(2020)159}{{\em JHEP} {\bfseries 06}
  (2020) 159}, \href{http://arxiv.org/abs/2003.07373}{{\ttfamily
  arXiv:2003.07373 [hep-th]}}.

\bibitem{Beratto:2020qyk}
E.~Beratto, N.~Mekareeya, and M.~Sacchi, ``{Marginal operators and
  supersymmetry enhancement in 3d $S$-fold SCFTs},''
  \href{http://dx.doi.org/10.1007/JHEP12(2020)017}{{\em JHEP} {\bfseries 12}
  (2020) 017}, \href{http://arxiv.org/abs/2009.10123}{{\ttfamily
  arXiv:2009.10123 [hep-th]}}.

\bibitem{Green:2010da}
D.~Green, Z.~Komargodski, N.~Seiberg, Y.~Tachikawa, and B.~Wecht, ``{Exactly
  Marginal Deformations and Global Symmetries},''
  \href{http://dx.doi.org/10.1007/JHEP06(2010)106}{{\em JHEP} {\bfseries 06}
  (2010) 106}, \href{http://arxiv.org/abs/1005.3546}{{\ttfamily arXiv:1005.3546
  [hep-th]}}.

\bibitem{Giambrone:2021zvp}
A.~Giambrone, E.~Malek, H.~Samtleben, and M.~Trigiante, ``{Global Properties of
  the Conformal Manifold for S-Fold Backgrounds},''
  \href{http://arxiv.org/abs/2103.10797}{{\ttfamily arXiv:2103.10797
  [hep-th]}}.

\bibitem{Gunaydin:1984qu}
M.~Gunaydin, L.~J. Romans, and N.~P. Warner, ``{Gauged N=8 Supergravity in
  Five-Dimensions},''
\href{http://dx.doi.org/10.1016/0370-2693(85)90361-2}{{\em Phys. Lett.}
  {\bfseries 154B} (1985) 268--274}.

\bibitem{Schwarz:1983qr}
J.~H. Schwarz, ``{Covariant Field Equations of Chiral N=2 D=10 Supergravity},''
\href{http://dx.doi.org/10.1016/0550-3213(83)90192-X}{{\em Nucl.Phys.}
  {\bfseries B226} (1983) 269}.

\bibitem{Howe:1983sra}
P.~S. Howe and P.~C. West, ``{The Complete N=2, D=10 Supergravity},''
\href{http://dx.doi.org/10.1016/0550-3213(84)90472-3}{{\em Nucl.Phys.}
  {\bfseries B238} (1984) 181}.

\bibitem{Bobev:2016nua}
N.~Bobev, H.~Elvang, U.~Kol, T.~Olson, and S.~S. Pufu, ``{Holography for $
  \mathcal{N} = 1^*$ on S$^{4}$},''
  \href{http://dx.doi.org/10.1007/JHEP10(2016)095}{{\em JHEP} {\bfseries 10}
  (2016) 095},
\href{http://arxiv.org/abs/1605.00656}{{\ttfamily arXiv:1605.00656 [hep-th]}}.

\bibitem{Andrade:2013gsa}
T.~Andrade and B.~Withers, ``{A simple holographic model of momentum
  relaxation},'' \href{http://dx.doi.org/10.1007/JHEP05(2014)101}{{\em JHEP}
  {\bfseries 1405} (2014) 101},
\href{http://arxiv.org/abs/1311.5157}{{\ttfamily arXiv:1311.5157 [hep-th]}}.

\bibitem{Mateos:2011tv}
D.~Mateos and D.~Trancanelli, ``{Thermodynamics and Instabilities of a Strongly
  Coupled Anisotropic Plasma},''
  \href{http://dx.doi.org/10.1007/JHEP07(2011)054}{{\em JHEP} {\bfseries 1107}
  (2011) 054},
\href{http://arxiv.org/abs/1106.1637}{{\ttfamily arXiv:1106.1637 [hep-th]}}.

\bibitem{Donos:2014eua}
A.~Donos and J.~P. Gauntlett, ``{Flowing from AdS$_{5}$ to AdS$_{3}$ with
  T$^{1,1}$},'' \href{http://dx.doi.org/10.1007/JHEP08(2014)006}{{\em JHEP}
  {\bfseries 08} (2014) 006}, \href{http://arxiv.org/abs/1404.7133}{{\ttfamily
  arXiv:1404.7133 [hep-th]}}.

\bibitem{Jain:2014vka}
S.~Jain, N.~Kundu, K.~Sen, A.~Sinha, and S.~P. Trivedi, ``{A Strongly Coupled
  Anisotropic Fluid From Dilaton Driven Holography},''
  \href{http://dx.doi.org/10.1007/JHEP01(2015)005}{{\em JHEP} {\bfseries 1501}
  (2015) 005},
\href{http://arxiv.org/abs/1406.4874}{{\ttfamily arXiv:1406.4874 [hep-th]}}.

\bibitem{Donos:2016zpf}
A.~Donos, J.~P. Gauntlett, and O.~Sosa-Rodriguez, ``{Anisotropic plasmas from
  axion and dilaton deformations},''
  \href{http://dx.doi.org/10.1007/JHEP11(2016)002}{{\em JHEP} {\bfseries 11}
  (2016) 002},
\href{http://arxiv.org/abs/1608.02970}{{\ttfamily arXiv:1608.02970 [hep-th]}}.

\bibitem{Donos:2013eha}
A.~Donos and J.~P. Gauntlett, ``{Holographic Q-lattices},''
  \href{http://dx.doi.org/10.1007/JHEP04(2014)040}{{\em JHEP} {\bfseries 1404}
  (2014) 040},
\href{http://arxiv.org/abs/1311.3292}{{\ttfamily arXiv:1311.3292 [hep-th]}}.

\bibitem{Freedman:1999gp}
D.~Z. Freedman, S.~S. Gubser, K.~Pilch, and N.~P. Warner, ``{Renormalization
  group flows from holography supersymmetry and a c theorem},''
  \href{http://dx.doi.org/10.4310/ATMP.1999.v3.n2.a7}{{\em Adv. Theor. Math.
  Phys.} {\bfseries 3} (1999) 363--417},
\href{http://arxiv.org/abs/hep-th/9904017}{{\ttfamily arXiv:hep-th/9904017
  [hep-th]}}.

\bibitem{Leigh:1995ep}
R.~G. Leigh and M.~J. Strassler, ``{Exactly marginal operators and duality in
  four-dimensional N=1 supersymmetric gauge theory},''
  \href{http://dx.doi.org/10.1016/0550-3213(95)00261-P}{{\em Nucl. Phys.}
  {\bfseries B447} (1995) 95--136},
\href{http://arxiv.org/abs/hep-th/9503121}{{\ttfamily arXiv:hep-th/9503121
  [hep-th]}}.

\bibitem{Maxfield:2016lok}
T.~Maxfield, ``{Supergravity Backgrounds for Four-Dimensional Maximally
  Supersymmetric Yang-Mills},''
  \href{http://dx.doi.org/10.1007/JHEP02(2017)065}{{\em JHEP} {\bfseries 02}
  (2017) 065},
\href{http://arxiv.org/abs/1609.05905}{{\ttfamily arXiv:1609.05905 [hep-th]}}.

\bibitem{Arav:2020obl}
I.~Arav, K.~C.~M. Cheung, J.~P. Gauntlett, M.~M. Roberts, and C.~Rosen,
  ``{Spatially modulated and supersymmetric mass deformations of $ \mathcal{N}
  $ = 4 SYM},'' \href{http://dx.doi.org/10.1007/JHEP11(2020)156}{{\em JHEP}
  {\bfseries 11} (2020) 156}, \href{http://arxiv.org/abs/2007.15095}{{\ttfamily
  arXiv:2007.15095 [hep-th]}}.

\bibitem{Guarino:2021kyp}
A.~Guarino and C.~Sterckx, ``{S-folds and holographic RG flows on the
  D3-brane},'' \href{http://arxiv.org/abs/2103.12652}{{\ttfamily
  arXiv:2103.12652 [hep-th]}}.

\bibitem{Bergshoeff:1980is}
E.~Bergshoeff, M.~de~Roo, and B.~de~Wit, ``{Extended Conformal Supergravity},''
\href{http://dx.doi.org/10.1016/0550-3213(81)90465-X}{{\em Nucl. Phys.}
  {\bfseries B182} (1981) 173--204}.

\bibitem{deRoo:1985np}
M.~de~Roo, ``{Gauged N=4 Matter Couplings},''
\href{http://dx.doi.org/10.1016/0370-2693(85)91619-3}{{\em Phys. Lett.}
  {\bfseries 156B} (1985) 331--334}.

\bibitem{deRoo:1984zyh}
M.~de~Roo, ``{Matter Coupling in N=4 Supergravity},''
\href{http://dx.doi.org/10.1016/0550-3213(85)90151-8}{{\em Nucl. Phys.}
  {\bfseries B255} (1985) 515--531}.

\bibitem{Bobev:2010de}
N.~Bobev, A.~Kundu, K.~Pilch, and N.~P. Warner, ``{Supersymmetric Charged
  Clouds in $AdS_5$},'' \href{http://dx.doi.org/10.1007/JHEP03(2011)070}{{\em
  JHEP} {\bfseries 03} (2011) 070},
\href{http://arxiv.org/abs/1005.3552}{{\ttfamily arXiv:1005.3552 [hep-th]}}.

\bibitem{Girardello:1999bd}
L.~Girardello, M.~Petrini, M.~Porrati, and A.~Zaffaroni, ``{The Supergravity
  dual of N=1 superYang-Mills theory},''
  \href{http://dx.doi.org/10.1016/S0550-3213(99)00764-6}{{\em Nucl. Phys.}
  {\bfseries B569} (2000) 451--469},
\href{http://arxiv.org/abs/hep-th/9909047}{{\ttfamily arXiv:hep-th/9909047
  [hep-th]}}.

\bibitem{Perlmutter:2020buo}
E.~Perlmutter, L.~Rastelli, C.~Vafa, and I.~Valenzuela, ``{A CFT Distance
  Conjecture},'' \href{http://arxiv.org/abs/2011.10040}{{\ttfamily
  arXiv:2011.10040 [hep-th]}}.

\bibitem{Lunin:2005jy}
O.~Lunin and J.~M. Maldacena, ``{Deforming field theories with U(1) x U(1)
  global symmetry and their gravity duals},''
  \href{http://dx.doi.org/10.1088/1126-6708/2005/05/033}{{\em JHEP} {\bfseries
  05} (2005) 033}, \href{http://arxiv.org/abs/hep-th/0502086}{{\ttfamily
  arXiv:hep-th/0502086}}.

\bibitem{Gauntlett:2005jb}
J.~P. Gauntlett, S.~Lee, T.~Mateos, and D.~Waldram, ``{Marginal deformations of
  field theories with AdS(4) duals},''
  \href{http://dx.doi.org/10.1088/1126-6708/2005/08/030}{{\em JHEP} {\bfseries
  08} (2005) 030}, \href{http://arxiv.org/abs/hep-th/0505207}{{\ttfamily
  arXiv:hep-th/0505207}}.

\bibitem{Bobev:2021gza}
N.~Bobev, P.~Bomans, F.~F. Gautason, and V.~S. Min, ``{Marginal deformations
  from type IIA supergravity},''
  \href{http://dx.doi.org/10.21468/SciPostPhys.10.6.140}{{\em SciPost Phys.}
  {\bfseries 10} (2021) 140}, \href{http://arxiv.org/abs/2103.02038}{{\ttfamily
  arXiv:2103.02038 [hep-th]}}.

\bibitem{Ashmore:2016oug}
A.~Ashmore, M.~Gabella, M.~Gra\~na, M.~Petrini, and D.~Waldram, ``{Exactly
  marginal deformations from exceptional generalised geometry},''
  \href{http://dx.doi.org/10.1007/JHEP01(2017)124}{{\em JHEP} {\bfseries 01}
  (2017) 124}, \href{http://arxiv.org/abs/1605.05730}{{\ttfamily
  arXiv:1605.05730 [hep-th]}}.

\bibitem{Ashmore:2018npi}
A.~Ashmore, ``{Marginal deformations of 3d $\mathcal{N}=2$ CFTs from AdS$_4$
  backgrounds in generalised geometry},''
  \href{http://dx.doi.org/10.1007/JHEP12(2018)060}{{\em JHEP} {\bfseries 12}
  (2018) 060}, \href{http://arxiv.org/abs/1809.03503}{{\ttfamily
  arXiv:1809.03503 [hep-th]}}.

\bibitem{Bobev:2021yya}
N.~Bobev, F.~F. Gautason, and J.~van Muiden, ``{The Holographic Conformal
  Manifold of 3d $\mathcal{N}=2$ $S$-fold SCFTs},''
  \href{http://arxiv.org/abs/2104.00977}{{\ttfamily arXiv:2104.00977
  [hep-th]}}.

\bibitem{Gunaydin:1985cu}
M.~Gunaydin, L.~J. Romans, and N.~P. Warner, ``{Compact and Noncompact Gauged
  Supergravity Theories in Five-Dimensions},''
\href{http://dx.doi.org/10.1016/0550-3213(86)90237-3}{{\em Nucl. Phys.}
  {\bfseries B272} (1986) 598--646}.

\bibitem{Khavaev:2000gb}
A.~Khavaev and N.~P. Warner, ``{A Class of N=1 supersymmetric RG flows from
  five-dimensional N=8 supergravity},''
  \href{http://dx.doi.org/10.1016/S0370-2693(00)01228-4}{{\em Phys. Lett.}
  {\bfseries B495} (2000) 215--222},
\href{http://arxiv.org/abs/hep-th/0009159}{{\ttfamily arXiv:hep-th/0009159
  [hep-th]}}.

\end{thebibliography}

\providecommand{\href}[2]{#2}\begingroup\raggedright\endgroup

\end{document}